\documentclass[12pt, reqno]{amsart}
\usepackage[a4paper, portrait, margin=25mm]{geometry}
\pdfoutput=1
\usepackage{latexsym,amssymb} 
\usepackage{mathtools} 
\usepackage{graphicx}
\graphicspath{ {./images/} } 
\usepackage[hidelinks]{hyperref}
\usepackage[font=small,labelfont=bf]{caption}
\usepackage{subcaption}
\usepackage{sidecap}
\usepackage{appendix}
\usepackage{algorithm} 
\usepackage{algpseudocode} 
\usepackage{epstopdf}
\usepackage{amsmath}
\usepackage[utf8]{inputenc}
\usepackage[T1]{fontenc}
\usepackage[scaled]{helvet}
\usepackage[foot]{amsaddr}
\usepackage{natbib}
\usepackage{xcolor}
\usepackage{color}
\usepackage{fancyhdr}
\usepackage{comment}
\usepackage{soul}
\usepackage{amsthm}
\newtheorem{definition}{Definition}

\newcommand{\defeq}{\vcentcolon=}

\title{\vspace*{-40pt}Recoverability of ancestral\\ recombination graph topologies}
\author{Elizabeth Hayman$^{1,*}$}
\address{$^1$ \textnormal{Department of Mathematics, University of Oxford, Andrew Wiles Building, Oxford OX2 6GG, UK}}
\author{Anastasia Ignatieva$^{2,3}$}
\address{$^2$ \textnormal{Department of Statistics, University of Warwick, Coventry CV4 7AL, UK}}
\author{Jotun Hein\,$^{3,4}$}
\address{$^3$ \textnormal{Department of Statistics, University of Oxford, 24-29 St Giles', Oxford OX1 3LB, UK}}
\address{$^4$ \textnormal{The Alan Turing Institute, British Library, London NW1 2DB, UK}}
\email{\ttfamily{elizabeth.hayman@keble.ox.ac.uk}}

\makeatletter
\renewcommand\@setemails{%
\mbox{{\itshape $^*$ E-mail}:\space}{\emails}
}
\makeatother

\date{\today}

\begin{document}

\maketitle


\vspace{-20pt}
\begin{abstract}

Recombination is a powerful evolutionary process that shapes the genetic diversity observed in the populations of many species. Reconstructing genealogies in the presence of recombination from sequencing data is a very challenging problem, as this relies on mutations having occurred on the correct lineages in order to detect the recombination and resolve the ordering of coalescence events in the local trees. We investigate the probability of reconstructing the true topology of ancestral recombination graphs (ARGs) under the coalescent with recombination and gene conversion. We explore how sample size and mutation rate affect the inherent uncertainty in reconstructed ARGs, which sheds light on the theoretical limitations of ARG reconstruction methods. We illustrate our results using estimates of evolutionary rates for several organisms; in particular, we find that for parameter values that are realistic for SARS-CoV-2, the probability of reconstructing genealogies that are close to the truth is low. 

\end{abstract}

\vspace{-15pt}
\renewcommand{\abstractname}{Keywords}
\begin{abstract}
Recombination detection, ancestral recombination graph, coalescent, gene conversion.
\end{abstract}

\section{Introduction}

The reconstruction of genealogies from sequencing data in the presence of recombination has remained an important but challenging problem. Several tools have been developed for recovering the topology of genealogies, with some recent methods capable of tackling very large datasets using heuristic and approximate approaches \citep[e.g.][]{Kelleher19, relate}. However, all methods that use sequencing data rely on mutations in the genealogical history in order to detect recombination and determine the ordering of coalescence events. Particularly when mutation rates are low, there may thus be significant uncertainty in the shape of the reconstructed local trees. Some tools \citep[such as ARGweaver,][]{argweaver} instead infer a distribution over genealogies, allowing inference methods to integrate over this uncertainty, although these are generally limited by computational power and can handle only moderate sample sizes.

In this article, we calculate the probability that the true topology of the genealogy (disregarding branch lengths) can be recovered from the data, either in full or up to a specified number of ambiguous internal edges, under some simplifying assumptions. This sheds light both on the performance of heuristic reconstruction methods (by quantifying how close to the true history they might get in the best case scenario) and methods exploring the posterior distribution over compatible genealogies (by giving a sense of the size of the search space). 

The coalescent with recombination is a widely used model for genealogies that extends coalescent trees to ancestral recombination graphs (ARGs) \citep{griffithsmarjoram}. Under the commonly used \emph{infinite sites} assumption, which we implement, each mutation occurs at a new position of the genome. Recombination can then be detected using the \emph{four gamete} test \citep{Hudson85}: denoting the ancestral allele by 0 and the derived allele by 1, if all four configurations 00, 01, 10 and 11 are observed at any two sites of a sample, then the sample could not have been generated by mutation alone and at least one recombination must have occurred. For a recombination to generate such incompatible sites, the ARG topology must include a particular configuration of coalescence events preceding a recombination, and mutations must fall on the correct edges of the recombination cycle. 

Under the coalescent with recombination, \citet{Myersthesis} derived the probability that, conditional on a single recombination having occurred in the history of a sample, its effect is detectable from the sequencing data. This was achieved by constructing recursion equations for the probability of interest, starting at the present time and considering each subsequent event backwards in time. We utilise similar ideas to consider the detectability of multiple recombination events, under the simplifying assumption of a two-locus model (where two non-recombining segments are separated by a single recombination breakpoint). We also make the assumption that the ARG topology is constrained to be a \emph{galled tree}, i.e.\ an ARG where the recombination cycles do not interact with each other. This allows us to calculate the probability that, conditioning on $R$ recombination events having occurred in the sample's history, they are all detectable, and the topology of each local tree can be reconstructed unambiguously (or up to a fixed number of ambiguous internal edges). We also calculate the probability that an ARG generated under the coalescent with recombination is a galled tree, and find that this is a reasonable assumption if the recombination rate is relatively low. We also consider gene conversion---where a section of genetic material is taken from one parent genome, and the endpoints from another parent genome---and derive the probability that given one gene conversion event has occurred in the history of the sample, this is detectable from the sequencing data.

The idea of constructing recursion equations to calculate quantities of interest for the coalescent with recombination goes back several decades: \citet{ethier1990two}, and later \citet{jenkins2009closed, jenkins2010asymptotic}, used recursion equations to obtain (asymptotic) closed-form expressions for the two-locus sampling distribution under different mutation models. Our work also links with previous explorations of the properties of coalescent genealogies \emph{without} recombination. In this setting, \citet{wiuf99} considered the age of a single mutation conditioned on its prevalence among sampled sequences; \citet{sargsyan2006analytical}, \citet{Hobolth2009TheGS}, \citet{jenkins11} and \citet{jenkins14} extended this to incorporate recurrent mutations under various conditions on the placement of mutations on the genealogy and on the demographic model. Our work explores a new direction in that we do not focus on sampling distributions or the properties of mutations such as the frequency spectrum directly; we instead use recursion equations to calculate the probability of mutations falling on the genealogy in a way that makes the ordering of coalescence events and presence of recombination events deducible from the data.

There is also a large body of literature concerned with assessing the effect of sample size and sequence length on the accuracy of phylogenetic inference, incorporating both simulation studies and theoretical investigations for a multitude of settings and mutation models \citep[e.g.][]{hillis1994hobgoblin, hillis1998taxonomic, kim1998large, pollock2002increased, heath2008taxon}. Our investigation of the accuracy of reconstructed tree topologies for data generated under the coalescent (without recombination) supports the broad consensus that accuracy improves with increasing the number of sampled taxa (when fixing the number of leaves for which the phylogeny is required) and sequence length. We derive analytic expressions for the probabilities of interest under the coalescent, which are new to the best of our knowledge; we also explicitly consider the presence of recombination, which has not been the focus of these prior studies.

Where possible, we illustrate our findings using mutation and recombination rate parameters that are reasonable for biological organisms. Using published estimates of evolutionary rates for SARS-CoV-2, we take the population scaled mutation and recombination rates to be approximately $\theta = 100$ and $\rho = 0.1$ per genome, respectively (assuming a generation time of 7.5 days \citep{li_gentime}, $N_e = 50$, mutation rate of $1 \cdot 10^{-3}$ per site per year \citep{Duchene20}, recombination rate of $2 \cdot 10^{-6}$ per site per year \citep{Muller21}). We also consider \textit{Drosophila melanogaster}, with $\theta = 8$ and $\rho = 21$ per kb, using estimates of \citet{Chan12}. For human populations, typical rates are $\theta = \rho = 0.1$ per kb, as used in previous analyses \citep{Kelleher19}.

In Section \ref{ch_trees}, we first demonstrate our ideas in the simpler case where recombination is disallowed, i.e.\ when the genealogy is constrained to be a binary tree generated under the coalescent model. In Section \ref{ch_galled_tree}, we calculate the probability that an ARG generated under the coalescent with recombination is a galled tree. Then, in Section \ref{ch_recs}, we consider the probability of reconstructing the ARG topology unambiguously when it is a galled tree. Further, in Section \ref{ch_geneconv}, we derive the probability that a gene conversion event is detectable from sequencing data. A discussion of these results is presented in Section \ref{ch_discussion}.

MATLAB code used for the numerical calculations can be found on GitHub at \url{github.com/Elizabeth-Hayman/Recoverability_of_ARG_topologies}.

\section{Topology of a tree} \label{ch_trees}

We first assume the absence of recombination, so the genealogy can be represented by a rooted binary tree, and consider the probability that the true tree topology can be reconstructed unambiguously from a sample of sequencing data generated under the coalescent model.

\begin{figure}[htbp!]
  \centering
    \includegraphics[scale=.75]{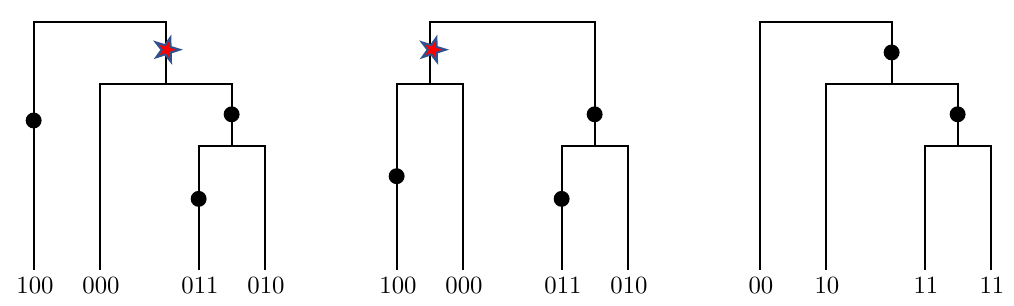}
    \caption{Examples of tree topologies that are and are not recoverable. Mutations are shown as dots, labelled by the site they affect; internal edges that do not carry a mutation are marked with red stars.}
   \label{fig:missed interior branches}
\end{figure}

Consider a rooted binary tree topology (disregarding branch lengths) as a directed acyclic graph (DAG), with a root node of in-degree 0 and out-degree 2, tree nodes with in-degree 1 and out-degree 2, and leaves with in-degree 1 and out-degree 0. The edges of this tree can then be classed as internal (connecting the root or a tree node with a tree node) and external (connecting the root or a tree node with a leaf). Each edge in the tree corresponds to a bipartition of the leaves, into those that are and those that are not descendants of the edge. A mutation on an edge manifests as a segregating site in the sequencing data, splitting the sampled sequences into those that do and those that do not carry the mutation, therefore allowing for the presence of the edge in the genealogy to be deduced. In fact, as a consequence of the splits-equivalence theorem \citep{buneman, semple2003phylogenetics}, if each internal edge undergoes at least one mutation, the tree topology can be uniquely determined from the sample. Note that mutations on the external edges are not helpful for resolving the ordering of coalescence events in the tree topology. This leads to the following definition:
\begin{definition}
A tree is \emph{recoverable} if at least one mutation is present on each of the internal edges, and hence the ordering of coalescence events can be uniquely determined from the sequencing data. 
\end{definition}
In Figure \ref{fig:missed interior branches}, the first two trees are not recoverable, and they are both consistent with the same sequencing sample. The rightmost tree is recoverable, as the topology is uniquely associated with the sample (even though fewer mutations occur overall).

Note that although we disregard recombination for now, some algorithms detect recombination by identifying changes to the local trees on either side of a recombination breakpoint \citep[]{Hein05}, so the detectability of recombination depends on how accurately the tree topologies at each locus can be reconstructed. 

\subsection{Probability that the tree is recoverable} \label{ch_s_full_top}

In order to derive the probability that the tree is recoverable, we proceed by considering the genealogy backwards in time, and tracking whether at least one mutation has occurred on each internal branch before it undergoes a coalescence event. At any point in time, each lineage can thus be in one of two states: if the lineage has not mutated since the last coalescence event, we assign it to being in \emph{State 1}, and otherwise in \emph{State 2}. Note that the terminal branches are taken to be in State 2 (mutations on the terminal branches are not required in order for the tree to be recoverable, as they provide no information on the ordering of coalescence events). 

At some point $t$ backwards in time, let ${n_l}$ be the total number of lineages currently remaining, and ${n_f}$ the number of lineages currently in State 2. Define $P_1(n_l, n_f)$ as the probability that the tree before (above) time $t$ is recoverable, given that $n_f$ out of $n_l$ lineages are in State 2 at $t$. Assigning all lineages to be in State 2 at the present time $t = 0$, $P_1(n, n)$ then gives the probability that the whole tree is recoverable.

We then construct recursion equations by conditioning on the current state pair $(n_f, n_l)$ and considering the possible next event backwards in time. Letting $\lambda = \binom{n_l}{2} + n_l \; \theta / 2$,
\begin{itemize}
    \item with probability $\binom{n_f}{2}/\lambda$ the event is a coalescence of two lineages in State 2 (then the number of lineages decreases by one, and the number of lineages in State 2 decreases by two);
    \item with probability $(n_l-n_f)\theta/2\lambda$ the event is a mutation of a lineage in State 1 (then the number of lineages in State 2 increases by one, and consequently the number of lineages in State 1 drops by one);
    \item with probability $n_f\theta/2\lambda$, the event is a mutation of a lineage in State 2 (then there is no change of state).
\end{itemize}
The recursion thus takes the form
 \begin{align} \label{full_top_known}
     \Bigg( \binom{n_l}{2} + n_l\frac{\theta}{2} \Bigg) P_1(n_l, n_f) = &\binom{n_f}{2} P_1(n_l-1, n_f-2) \\ \nonumber
     &+ (n_l-n_f) \frac{\theta}{2} P_1(n_l, n_f+1) \\ \nonumber
     &+ n_f \; \frac{\theta}{2} P_1(n_l, n_f), \;\;\; 0 \leq n_f \leq n_l,
 \end{align} 
 with the initial condition $P_1(1, 0)=1$. This is the simplest case of recursions we present, which can be solved numerically with a runtime of order roughly $n^2$. Further on in the text, recursions become more complex, but can be solved efficiently via dynamic programming and should not require any matrix inversion. We use MATLAB to solve the recursions.

\begin{figure}[htbp!]
  \centering
  \begin{subfigure}[t]{0.45\linewidth}
    \includegraphics[trim={4cm 9.2cm 4cm 9.2cm},clip,width=\linewidth]{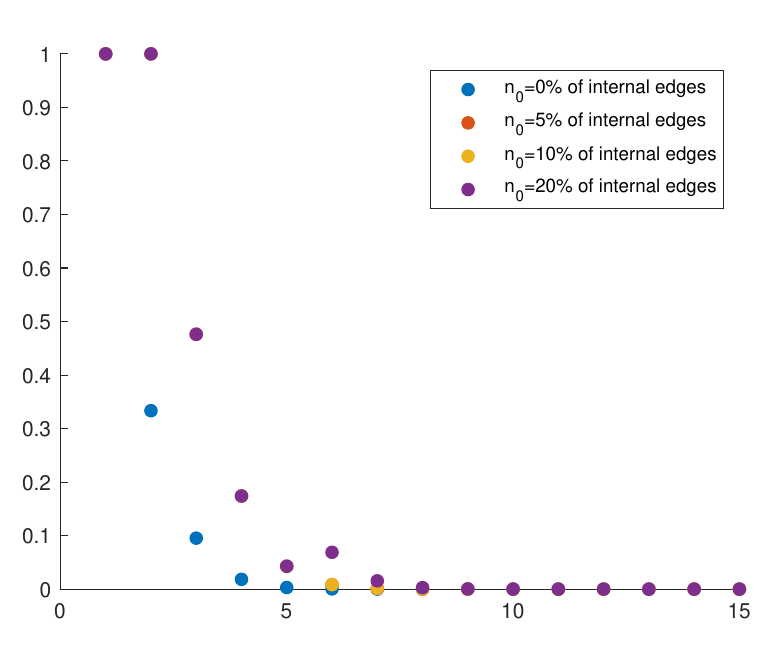}
    \caption{$\theta=1$.}
  \end{subfigure}
  \quad
  \begin{subfigure}[t]{0.45\linewidth}
    \includegraphics[trim={4cm 9.2cm 4cm 9.2cm}, clip,width=\linewidth]{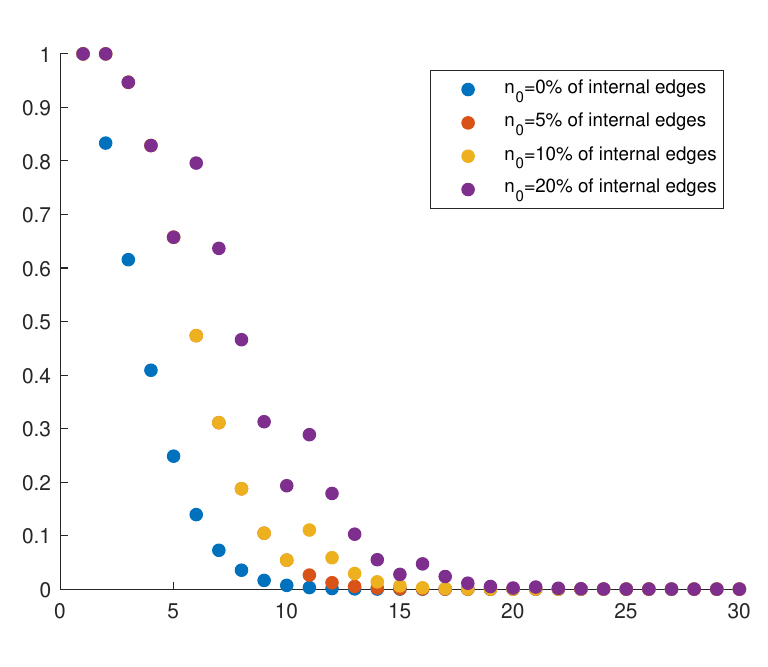}
    \caption{$\theta=10$.}
  \end{subfigure}
  \quad
  \begin{subfigure}[t]{0.45\linewidth}
    \includegraphics[trim={4cm 9.2cm 4cm 9.2cm}, clip,width=\linewidth]{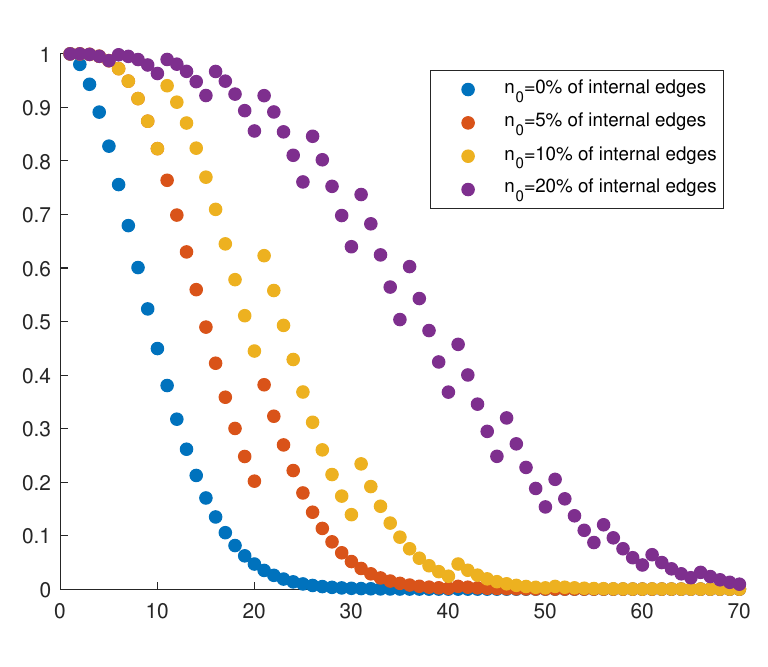}
    \caption{$\theta=100$.}
  \end{subfigure}
  \quad
  \begin{subfigure}[t]{0.45\linewidth}
    \includegraphics[trim={4cm 9.2cm 4cm 9.2cm}, clip,width=\linewidth]{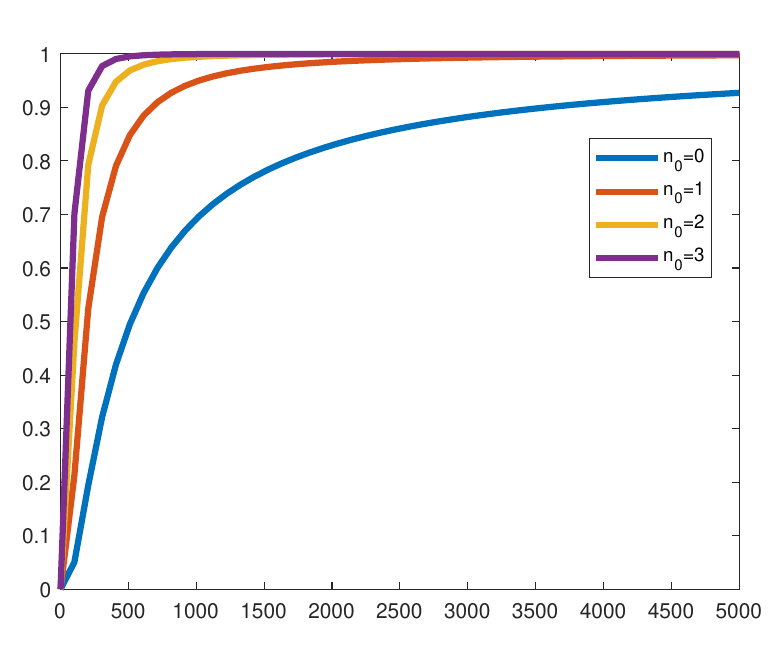}
    \caption{For a fixed sample $n=20$ varying $\theta$ over biologically feasible values.}
  \end{subfigure}
  \caption{Probability of tree being (partially) recoverable. Panels (a)-(c) show the probabilities for different values of the mutation rate against sample size, for a range of $n_0$ (colours) taken to be a given percentage of the total number of internal edges in the tree (rounded down to the nearest integer). Panel (d) shows the probabilities for a fixed sample size $n=20$ against $\theta$, for a range of $n_0$ (colours).}
  \label{fig:_skipped_mutations_graphs}
\end{figure}

Figure \ref{fig:_skipped_mutations_graphs} (blue dots and lines) illustrates the probability of the tree being recoverable for various values of $n$ and $\theta$. Panels (a)-(c) demonstrate that this probability is monotonically decreasing in $n$ for fixed $\theta$; for larger values of $n$, the time periods between coalescence events are shorter near the present time, so it is less likely that mutations will occur on all of the internal edges. Panel (d) shows that the probability that the tree is recoverable for a sample of fixed size ($n = 20$) increases as $\theta$ grows, with a limit of 1 as $\theta \rightarrow \infty$, as expected. However, the probability that the tree is recoverable does not reach near 1 until $\theta \approx 10^5$, which appears infeasibly large for biological samples (typically $\theta \leq 100$). For $\theta = 100$, which is typical for SARS-CoV-2 genomes for instance, the probability of the tree being recoverable becomes very small for sample sizes over 25. For \emph{Drosophila melanogaster}, with $\theta \approx 10$, this probability is minuscule for $n > 10$.
 

\citet{Fu693} derive the expected total length of internal branches to be 
\[
2 \left((\sum_{j=1}^{n-1} 1/j) - 1 \right) \approx \log(n)
\] 
for large $n$, using our time scaling. Given that mutations occur as a Poisson process with rate $\theta/2$, the expected total number of mutations on interior branches is therefore $\approx \theta \log(n)$. This gives some intuitive understanding of the results above: for a sample of size $n$, there are $n-1$ coalescent events, and thus a minimum of $n-2$ mutations are required to have one on each interior branch. Therefore, even before the precise placement of individual mutations is considered, the total number of mutations needed for the tree to be recoverable grows like $n$, while the number of mutations expected to occur on the interior branches grows like $\log(n)$. Hence, the probability of the tree being recoverable drops to 0 quickly.

\subsection{Probability that tree is partially recoverable}

The probability of the tree being recoverable decreases rapidly as $n$ increases, so we next consider the probability of the tree being \emph{partially} recoverable: 
\begin{definition}
A tree is \emph{partially recoverable} if all except up to $n_0$ internal edges carry at least one mutation. 
\end{definition}
Figure \ref{fig:missed interior branches} (left and centre panels) demonstrates two possible trees for $n=4$ and $n_0=1$: the trees are consistent with the same sample, and have the same topology apart from the internal edge joining the second sequence to the rest of the tree. A mutation at one of the starred positions would fix the reconstructed genealogy to one of these two possibilities.

We extend the results of the previous section by including $n_0$ as a recursive index to track the number of internal edges that have not yet undergone at least one mutation. By again considering the next event backwards in time when there are $n_l$ lineages of which $n_f$ are in State 2, the equivalent recursion to \eqref{full_top_known} is
\begin{align*}
   \left( \binom{n_l}{2} + (n_l-n_f)\frac{\theta}{2} \right) P_{1a}(n_0,n_l,n_f) = &\binom{n_f}{2} P_{1a}(n_0,n_l-1,n_f-2)\\
   &+ (n_l-n_f) \frac{\theta}{2} P_{1a}(n_0,n_l,n_f+1) \\ 
   &+ n_f(n_l-n_f) P_{1a}(n_0-1,n_l-1,n_f-1) \\
   &+ \binom{n_l-n_f}{2}P_{1a}(n_0-2,n_l-1,n_f).
\end{align*}
The initial conditions are $P_{1a}(n_0, 1, 1)=1=P_{1a}(n_0, 1, 0)$. Note that we are considering the probability of having \emph{up to} $n_0$ unresolved internal edges: the probability of missing \emph{precisely} $n_0$ edges can be calculated as $P_{1a}(n_0, 1, 1)-P_{1a}(n_0-1, 1, 1)$.

Figure \ref{fig:_skipped_mutations_graphs} (a)-(c) shows how the probability of the tree being partially recoverable varies with $\theta$ and $n_0$, where $n_0$ is taken to be a fixed percentage of the total number of internal edges in the tree (rounded down to the nearest integer, which makes the plots look jagged). Note that increasing $n_0$ in panels (a)-(c) shifts the probability curve to the right, for each value of $\theta$, and that the magnitude of the shift increases with $\theta$. Panel (d) highlights that allowing even a small number of unresolved edges results in a large increase in the probability of correctly reconstructing the rest of the tree.

 \subsection{History of a specific lineage}
\begin{figure}[htbp!]
  \centering
    \includegraphics[scale=.8]{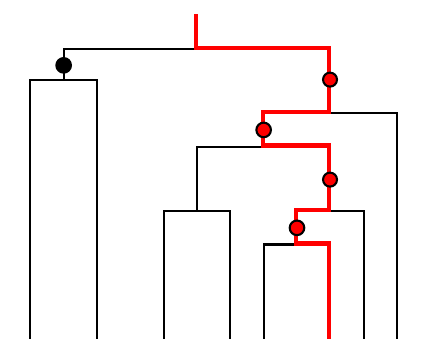}
    \caption{Example of a tree with a single lineage of interest (highlighted in red).}
   \label{fig:example_single_lineage}
\end{figure}

The probability of the tree being even partially recoverable decreases rapidly with increasing sample size. We next focus on the probability of unambiguously determining the history of a specific lineage. This is useful if there is particular interest in the history of a specific sequence: for instance, in the context of viral genealogies, our results quantify the probability that a particular viral strain can be accurately placed in the overall genealogy. Figure \ref{fig:example_single_lineage} shows an example of a tree topology where the history of the lineage highlighted in red can be reconstructed unambiguously, while allowing for uncertainty in the rest of the tree topology. This requires that mutations occur on all of the internal edges highlighted in red.

This simplifies the model considered in Section \ref{ch_s_full_top}, as the dependence on $n_f$ can be dropped, with the recursions only focussing on the state of the one lineage. Here, ${n_0}$ counts only those unresolved internal edges that are ancestral to the single lineage of interest. The equivalent to \eqref{full_top_known} in this setting with $P_2(n_0, n_l, \text{lineage state})$ as the probability of interest is
 \begin{align}
     \left(\binom{n_l}{2}+\frac{\theta}{2}\right) P_2(n_0, n_l, 1) &= \binom{n_l-1}{2} P_2(n_0, n_l-1, 1) + (n_l-1) P_2(n_0-1, n_l-1, 1)\nonumber \\ 
     &+ \frac{\theta}{2} P_2(n_0, n_l, 2), \nonumber \\
     \binom{n_l}{2} P_2(n_0, n_l, 2) &= \binom{n_l-1}{2} P_2(n_0, n_l-1, 2) + (n_l-1) P_2(n_0, n_l-1, 1). 
 \end{align}
The initial conditions are $P_2(n_0, 1, 1)=1 = P_2(n_0, 1, 2)$.

\begin{figure}[htbp!]
  \centering
  \begin{subfigure}[t]{0.45\linewidth}
    \includegraphics[trim={0 0 0 .72cm},clip,width=\linewidth]{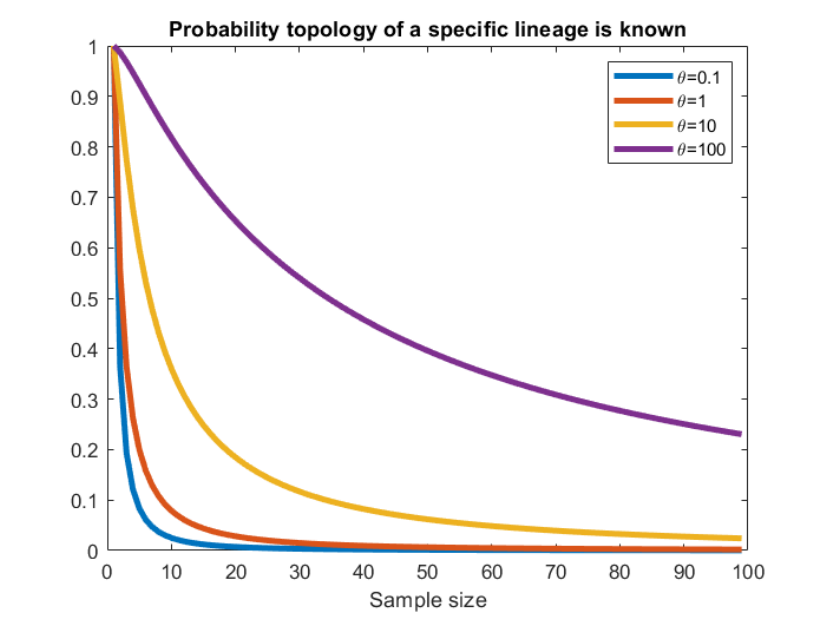}
    \caption{Varying $n$ for several fixed values of $\theta$ (colours).}
  \end{subfigure}
  \begin{subfigure}[t]{0.45\linewidth}
    \includegraphics[trim={0 0 0 .72cm},clip,width=\linewidth]{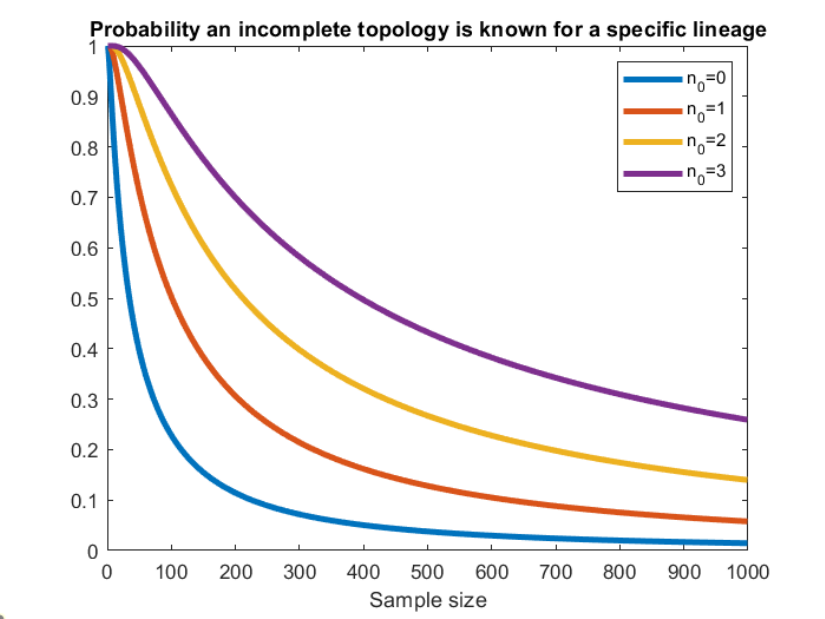}
    \caption{$\theta=100$, allowing up to $n_0$ undetermined internal edges (colours).}
  \end{subfigure}
  \caption{Probability of unambiguously reconstructing the history of a single lineage (left panel: unambiguously, right panel: with up to $n_0$ unresolved internal branches). Note the longer scale on the $x$-axis in panel (b).}
  \label{fig:one_lin_topology}
\end{figure}

Figure \ref{fig:one_lin_topology} shows plots of the resulting probabilities. The probabilities are consistently higher compared to those of the whole tree being recoverable, and are non-negligible even for reasonably large sample sizes. Taking the SARS-CoV-2 value of $\theta \approx 100$, panel (a) shows that the genealogical history of a particular lineage from a sample size of 20 has probability of $0.75$ of being reconstructed unambiguously. Panel (b) shows the probability of reconstructing the history of a single lineage up to $n_0$ unresolved internal edges. As before, even a small degree of flexibility (up to three undetermined interior edges out of a sample of hundreds) leads to a significant improvement in recoverability.

\section{Probability that an ARG is a galled tree} \label{ch_galled_tree}

\begin{figure}[htbp!]
  \centering
    \includegraphics[trim={0cm 10cm 0cm 9cm},clip,scale=.6]{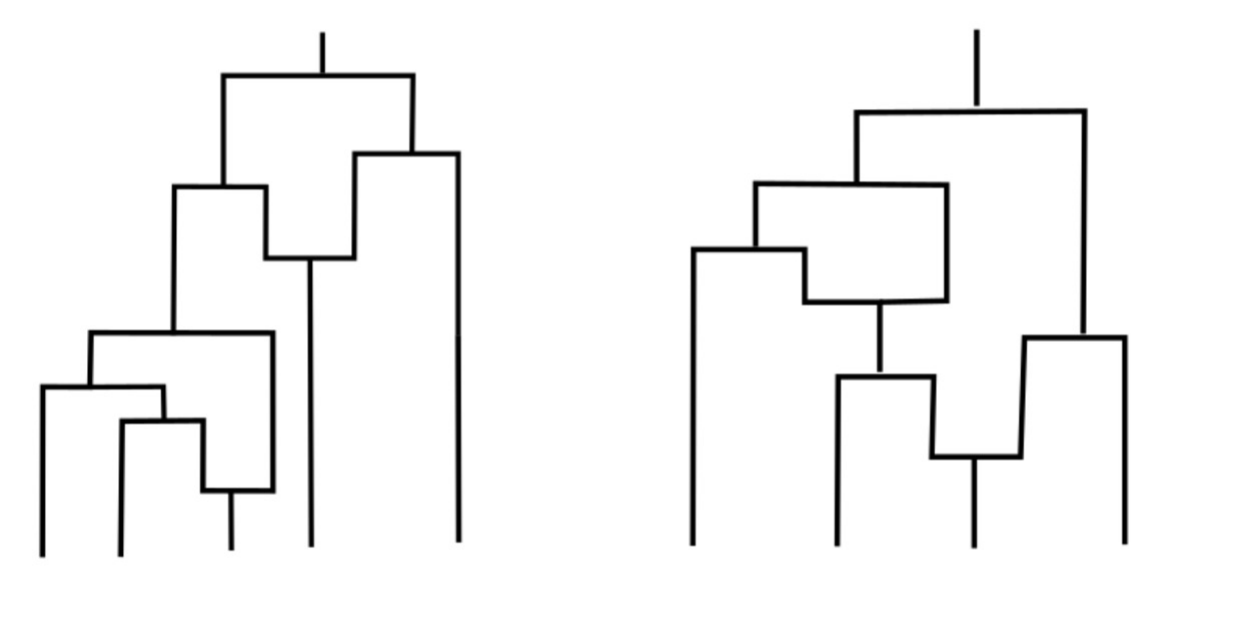}
    \caption{Left panel: ARG with two recombinations which is a galled tree (the recombination loops do not interact). Right panel: ARG that is not a galled tree, as the two recombinations loops are intertwined.}
  \label{fig:configurations_of_2_recombinations}
 \end{figure}

Define an ARG as a rooted binary DAG, containing a root node of in-degree 0 and out-degree 2, sample nodes with in-degree 1 and out-degree 0, recombination nodes with in-degree 2 and out-degree 1, and tree nodes with in-degree 1 and out-degree 2. Recombination nodes are labelled with a recombination breakpoint $z$ (which, assuming a two-locus model, is fixed), with the leftmost parent node inheriting the genetic material at coordinates $[0, z)$ and the rightmost parent inheriting the genetic material at $[z, 1]$. Suppose that two directed paths out of a node $x$ in the ARG meet at a recombination node $y$; a recombination \emph{loop} is the subgraph of the ARG containing $x$, $y$ and the two paths connecting them. If a recombination loop shares no node with any other recombination loop, it is termed \emph{galled} \citep[][p.\ 237]{Gusfield08}. A \emph{galled tree} is an ARG where all recombination loops are galled. An example of an ARG that is (resp.\ is not) a galled tree is given in the left (resp.\ right) panel of Figure \ref{fig:configurations_of_2_recombinations}.

Galled trees are a particularly tractable class of ARGs, for which many properties are significantly more straightforward to derive analytically, and reconstruction algorithms are attractively efficient. For instance, there exists a polynomial time algorithm for reconstructing a parsimonious galled tree from sequencing data, if this is possible \citep{wang01,gusfield04}; there is a concise necessary and sufficient condition for the sample to be consistent with a galled tree \citep{songgalledtrees}; if a genealogy is in the shape of a galled tree, the sample can also be derived on a true tree (with no recombination) if at most one recurrent mutation per site is allowed \citep[Theorem 8.12.1]{Gusfield08}.

\citet{Gusfield08} notes that ARGs are likely to be galled trees if the recombination rate is low, or if there is reason to believe that recombination has only occurred relatively close to the present. However, the probability that an ARG generated under the coalescent with recombination is a galled tree has not been previously derived analytically. We obtain an explicit expression for the probability that an ARG with $n$ leaves and known recombination rate $\rho$ contains only galled recombination cycles. Note that the results we derive hold for the \emph{big} ARG (i.e.\ including recombination events in non-ancestral material); this simplifies the expressions, and we expect the results to be very close to those for the \emph{small} ARG (ignoring recombination events in non-ancestral material) when $\rho$ is small and $n$ reasonably large.

Define an \emph{open} recombination loop as one where (looking backwards in time) the two recombinant lineages have not yet coalesced back with each other. An ARG may fail to be a galled tree if a lineage of an open recombination loop undergoes another recombination, or coalesces with a lineage from a different open recombination loop. 

\subsection{Probability that an ARG has exactly \texorpdfstring{$R$}{R} recombination nodes}

First, we obtain an expression that an ARG has precisely $R$ recombination events for a sample of size $n$. Consider the ARG some time $t$ before the present. Define $Q_1^{\rho, R}(n_l,r)$ as the probability that an ARG has at most $R$ recombinations (in total), given that there are ${n_l}$ lineages remaining at $t$, with $r \leq R$ possible recombination events having occurred before (below) time $t$. By considering the genealogy backwards in time and conditioning on the next possible event, we construct the following recursion for $Q_1^{\rho, R}(n_l,r)$:
\begin{align}
Q_1^{\rho, R}(n_l,r) = &\frac{n_l-1}{n_l-1+\rho} \cdot Q_1^{\rho, R}(n_l-1, r) \\
&+ \frac{\rho}{n_l-1+\rho} \cdot Q_1^{\rho, R}(n_l+1,r+1) \; \text{ for } r \leq R, \; n_l \leq n + R,\\
Q_1^{\rho, R}(n_l,r) &= 0 \; \text{ otherwise}.
\nonumber
\end{align}
The condition $n_l \leq n + R$ arises as at most $n+R$ lineages can be present in the ARG at any time (through the sample undergoing $R$ recombination before any coalescences). The initial condition is $Q_1^{\rho, R}(1, R)=1$, as the process must terminate with precisely $R$ recombinations having occurred. Then $\mathcal{Q}_1^{n, \rho, R} \defeq Q_1^{\rho, R}(n,0)$ gives the probability that the ARG has exactly $R$ recombination nodes, for a sample of size $n$.

\subsection{Probability that an ARG with \texorpdfstring{$R$}{R} recombination nodes is a galled tree}

We next derive the probability that an ARG with exactly $R$ recombination nodes is a galled tree. Again considering the ARG at some point $t$ backwards in time, let $Q_2^{\rho, R}(n_l, r_0, r)$ be the probability that the ARG has precisely $R$ recombination loops in total, all of which are galled, conditional on there being $n_l$ lineages at time $t$, with $r$ out of $R$ recombinations having occurred before $t$, and $r_0 \leq r$ recombination loops currently open. As illustrated in Figure \ref{fig:configurations_of_2_recombinations}, an ARG may fail to be a galled tree if (1) a lineage that is part of an open recombination loop undergoes a further recombination, or if (2) two lineages that are part of different open recombination loops coalesce. Considering each possible next event that does not lead to the ARG failing to be a galled tree, the following recursions on $Q_2^{\rho, R}(n_l, r_0, r)$ can therefore be constructed:
\begin{align}
    \frac{n_l}{2}&(n_l-1+\rho)Q_2^{\rho, R}(n_l, r_0, r) =r_0 \cdot Q_2^{\rho, R}(n_l-1, r_0-1, r) \nonumber \\
    &+\left(\frac{1}{2}(n_l-2r_0)(n_l-2r_0-1)+2r_0(n_l-2r_0)\right)Q_2^{\rho, R}(n_l-1, r_0, r) \nonumber \\
    &+\frac{\rho}{2}(n_l-2r_0)Q_2^{\rho, R}(n_l+1, r_0+1, r+1), \; \text{ for } r_0 \leq r \leq R, \; 2r_0 \leq n_l \leq n + r_0, \\
    &Q_2^{\rho, R}(n_l,r_0, r)  = 0 \; \text{ otherwise}. \nonumber
\end{align}
The condition $2 r_0 \leq n_l$ arises as there must be at least $2r_0$ lineages in the ARG when $r_0$ galled recombination loops are open; the condition $n_l \leq n + r_0$ arises as there can be at most $n + r_0$ lineages in the ARG at any time (through the sample undergoing $r_0$ recombinations before any coalescences). The initial condition is $Q_2^{\rho, R}(1,0,R) = 1$. Then $\mathcal{Q}_2^{n, \rho, R} \defeq Q_2^{\rho, R}(n,0,0)$ is the probability that the ARG has exactly $R$ recombination loops, all of which are galled.

\subsection{Probability that an ARG is a galled tree}

Combining the results above, $\mathcal{S}^{n, \rho, R} \defeq \mathcal{Q}_2^{n, \rho, R}/\mathcal{Q}_1^{n, \rho, R}$ gives the probability that an ARG with exactly $R$ recombination nodes and $n$ leaves is a galled tree. Then
\begin{equation*}
    \mathcal{S}^{n, \rho, \leq R} = \frac{\sum_{i=0}^{R}\mathcal{Q}_2^{n, \rho, i}}{\sum_{i=0}^{R}\mathcal{Q}_1^{n, \rho, i}}
\end{equation*} 
gives the probability that an ARG is a galled tree, conditional on \emph{up to} $R$ recombinations having occurred. Taking the limit $\mathcal{S}^{n, \rho} \defeq \lim_{R \to \infty} S^{n, \rho, \leq R}$ removes the conditioning on $R$ (as a finite number of recombinations occurs in any history with probability one), thus giving an unconditional probability that an ARG with $n$ leaves is a galled tree.

\begin{figure}[htbp!]
  \centering
  \begin{subfigure}[t]{0.45\linewidth}
    \includegraphics[trim={0 0 0 .5cm},clip,width=\linewidth]{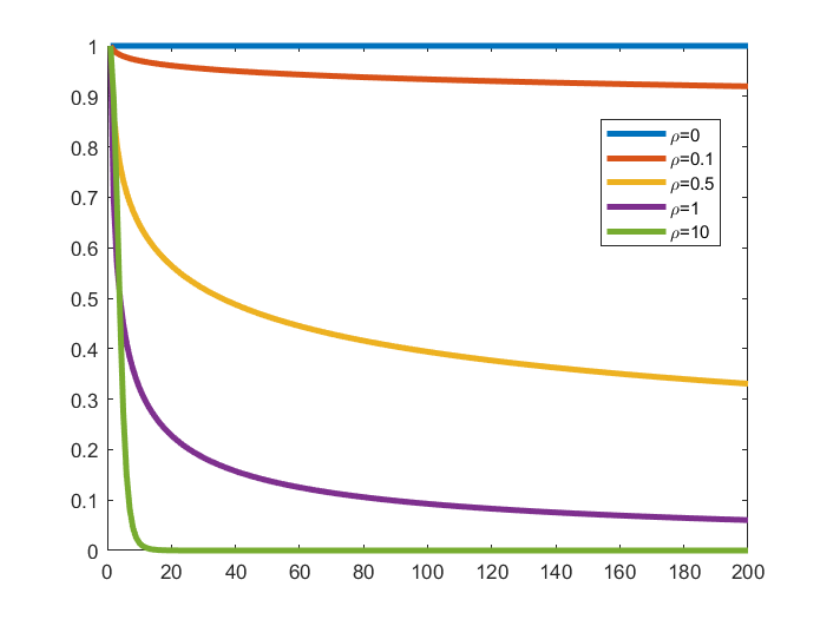}
    \caption{$\mathcal{S}^{n, \rho}$, varying $n$ for several fixed values of $\rho$ (colours).}
  \end{subfigure}
  \quad
  \vspace{10pt}
  \begin{subfigure}[t]{0.45\linewidth}
    \includegraphics[trim={0 0 0 .5cm},clip,width=\linewidth]{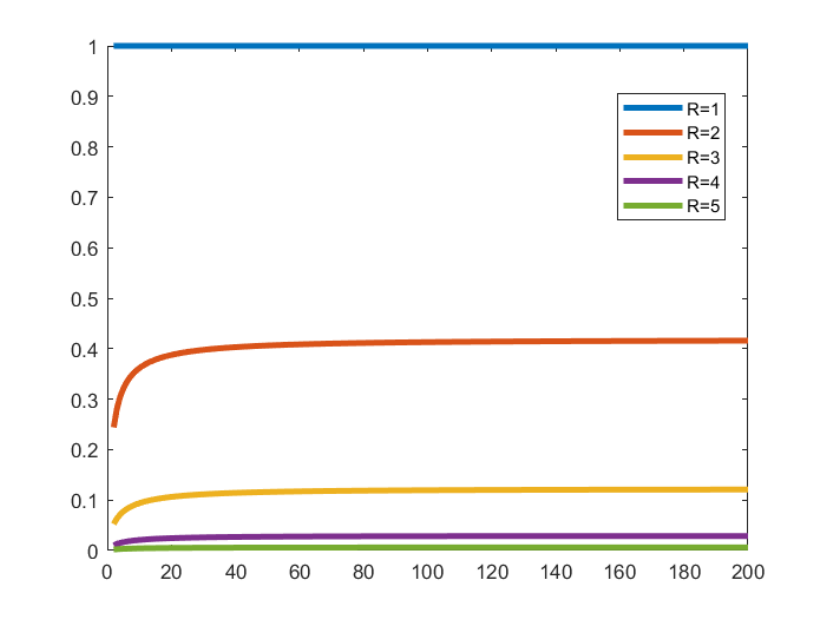} 
    \caption{$\mathcal{S}^{n, R}$, varying $n$, conditioning on the number of recombinations $R$ (colours) and averaging over $\rho \in (0,1)$ using a uniform prior.}
  \end{subfigure}
  \caption{Probability that an ARG is a galled tree for varying parameter values and sample size. }
  \label{fig:is_tree_galled?}
\end{figure}

The left panel of Figure \ref{fig:is_tree_galled?} illustrates the probability $\mathcal{S}^{n, \rho}$ that an ARG is a galled tree for a range of sample sizes $n$ and recombination rates $\rho$. When the recombination rate is low, ARGs are galled trees with high probability; this is both due to the ARGs being likely to contain at most one recombination node (and hence being trivially galled), or the recombinations being `far apart' in the ARG so that the recombination loops are not likely to interact. 

The right panel of Figure \ref{fig:is_tree_galled?} shows the probability $\mathcal{S}^{n, R} \defeq \int_0^1 q(\rho) \mathcal{S}^{n, \rho, R} d\rho$, integrating over a uniform prior distribution $q$ on the recombination rate $\rho \in (0,1)$, to illustrate the probability that the ARG is a galled tree conditioning on $R$ recombinations and assuming a low recombination rate. For $R=2$ recombinations, the ARG is a galled tree with reasonably high probability, of around 0.4 when the sample size is moderate. This suggests that restricting our consideration to ARGs in the form of galled trees might be reasonable when analysing whole-genome SARS-CoV-2 data, for instance, and human or drosophila samples of relatively short genomic regions.

\section{Topology of an ARG} \label{ch_recs}

We now extend our results in Section \ref{ch_trees} to include crossover recombination, through analysing the probability of reconstructing the ARG topology in the form of a galled tree unambigiously (or up to a specified number of ambiguous internal edges), under a two-locus model.

\subsection{Detectability of one recombination} \label{ch_s_myers}

Conditioning on exactly one recombination having occurred in the history of a sample with a breakpoint $z \in [0,1]$, \citet{Myersthesis} considers the probability of this recombination being \emph{detectable}:

\begin{definition}
A recombination is \emph{detectable} if it changes the ARG topology (i.e.\ the topology of at least one local tree), and mutations fall on the correct edges of the recombination loop to create incompatibilities in the data, which can then be detected by the four gamete test.
\end{definition}

\begin{figure}[htbp!]
  \centering
    \includegraphics[scale=0.9]{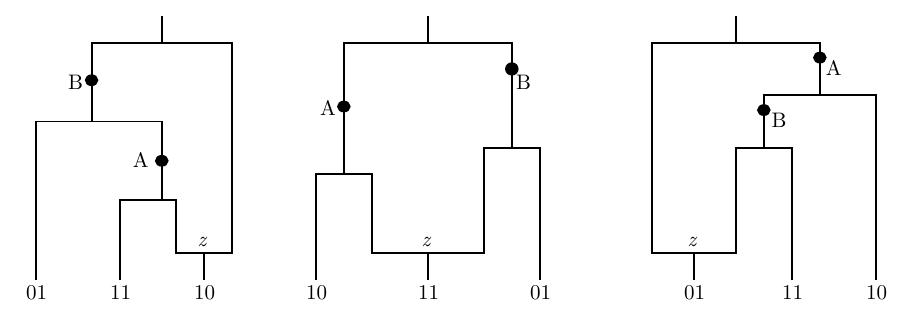}
    \caption{Positioning of mutations on the ARG with a single recombination that are required for the recombination to be detectable. Ancestral type is assumed to be $00$.}
  \label{fig:configurations of mutations leading to detectable recombinations}
 \end{figure} 
 
The necessary conditions on the ARG topology and positions of mutations are illustrated in Figure \ref{fig:configurations of mutations leading to detectable recombinations}. One of the shows configurations must be a subgraph of the ARG for the recombination to be detectable from the data. Labelling the locus to the left (resp.~right) of the breakpoint as $A$ (resp.~$B$), denote $A$-type mutations as those occurring in locus $A$, and $B$-type those in locus $B$. Assuming that the ancestral type is known to be $00$, all of the configurations shown in the Figure generate incompatible sites at $A$ and $B$.

\citet[Section 4.3]{Myersthesis} calculates the probability of the recombination being detectable through constructing recursion equations, beginning at the recombination event and tracking the state of each recombinant lineage backwards in time (dissallowing any further recombintions). The possible states for the recombinant lineage emerging from the left-hand side of the recombination node (denoted $\mathcal{E}$) are given in Table \ref{tab:recombstates1}. The states for the right-hand lineage (denoted $\mathcal{F}$) are equivalent, but with $A$ and $B$ reversed. Myers notes that for one recombination to be detectable, it is sufficient to have either lineage reach State 4, or for both lineages to simultaneously be in state $\geq2$.

\begin{table}[htbp!]
  \begin{center}
    \caption{States described for the left recombinant edge denoted $\mathcal{E}$}
    \label{tab:recombstates1}
    \begin{tabular}{l|l} 
      
      State 0 & No coalescence has occurred on edge $\mathcal{E}$ since the recombination. \\
      \hline
      State 1 & There has been at least one coalescence on edge $\mathcal{E}$ since\\
       & the recombination. No mutations have occurred on edge $\mathcal{E}$ since \\
       & the last coalescence.\\
       \hline
      State 2 & $\mathcal{E}$ has reached state 1 and a type $A$ mutation has occurred since\\
      &  the last coalescence. \\
      \hline
      State 3 & $\mathcal{E}$ has reached state 2 and undergone one further coalescence.\\
      \hline
      State 4 & $\mathcal{E}$ has reached state 3 and a type $B$ mutation has occurred since\\
      & the last coalescence. \\
      \end{tabular}
  \end{center}
\end{table}

Our work extends these results by considering more than one (galled) recombination, and calculating the probability that each recombination is detectable, and also that each local tree is (partially) recoverable. Note that when conditioning on $R$ recombination events, we again consider the \emph{big} ARG, and condition on $R$ recombination events in the entire history of the sample (including those that might occur in non-ancestral material). This substantially simplifies calculations, and we expect the difference with conditioning on $R$ recombination events in ancestral material to be negligible for small $\rho$ and reasonably large $n$.

\subsection{Probability that ARG is recoverable}
We arrive at the following definition:
\begin{definition}
The ARG is \emph{recoverable} if all recombinations are detectable and each local tree is recoverable (i.e.\ each edge of the ARG which is internal in at least one local tree has at least one mutation).
\end{definition}
We now calculate the probability that the ARG is a galled tree and recoverable, conditioning on $R$ recombinations having occurred. This requires at least one mutation on each edge which is internal to at least one local tree, and the correct sequence of coalescent and mutation events within each recombination loop (to ensure the presence of incompatible sites in the data). This necessitates incorporating more states into the recursions of Section \ref{ch_trees} to track each of the recombinant lineages. Note that the restriction to galled trees means that for each recombination to the detectable, the detection conditions stated above must hold (independently) for each recombination loop.

A fixed number of recombinations $R$ are allowed to occur in the history of a sample of size $n$. Suppose that at some point in time, the total number of remaining lineages is $n_l$, and the number of non-recombinant lineages which have undergone at least one mutation since the last coalescence event is $n_f$. The other indices are given in the third column of Table \ref{tab:recombstates}, tracking the number of left recombinant lineages in various states (with equivalent states for the right recombinant lineages). The indices $a,b,c_1,c_2,d_1,d_2,e_1,e_2$ (resp.~$i,j,k_1,k_2,l_1,l_2,m_1,m_2$) count the number of left (resp.~right) lineages in states 0, 1,..., 7. The index $r$ tracks the number of recombination events which have occurred, with $r_0$ recombination loops currently remaining open. Note that we must have 
\[
r = a+b+c_1+c_2+d_1+d_2+e_1+e_2 = i+j+k_1+k_2+l_1+l_2+m_1+m_2.
\]

\begin{table}[htbp!]
  \begin{center}
    \caption{States are described for the left recombinant edge denoted $\mathcal{E}$ (third column gives the index that counts the number of lineages in each state).}
    \label{tab:recombstates}
    \begin{tabular}{l|l|c} 
      
      State 0 & No coalescence has occurred on edge $\mathcal{E}$ since the recombination. & $a$\\
      \hline
      State 1 & There has been at least one coalescence since the recombination. & $b$\\
       & No mutations have occurred since the last coalescence.& \\
       \hline
      State 2 & $\mathcal{E}$ has reached state 1 and mutations \textit{not} including a type A  &$c_2$ \\
      &  mutation have occurred since the last coalescence. & \\
      \hline
      State 3 & $\mathcal{E}$ has reached state 1 and mutations including a type A  &$c_1$ \\
      &  mutation have occurred since the last coalescence. & \\
      \hline
      State 4 & $\mathcal{E}$ has reached state 3 and undergone one further coalescence. & $d_1$\\
      \hline
      State 5 & $\mathcal{E}$ has reached state 4 and mutations \textit{not} including a type B  &$e_2$ \\
      &  mutation have occurred since the last coalescence. & \\
      \hline
      State 6 & $\mathcal{E}$ has reached state 4 and mutations including a type B  &$e_1$ \\
      &  mutation have occurred since the last coalescence. & \\
      \hline
      State 7 & $\mathcal{E}$ has reached state 6 and undergone one further coalescence. & $d_2$\\
    \end{tabular}
  \end{center}
\end{table}

\begin{figure}[htbp!]
  \centering
    \includegraphics[width=0.9\textwidth]{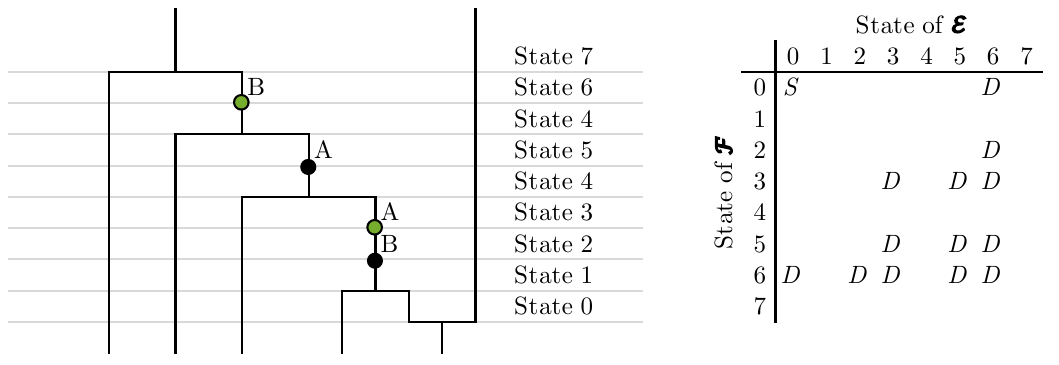}
    \caption{Progression of the recombinant lineage $\mathcal{E}$ through the states described in table \ref{tab:recombstates}. Note the mutations coloured green, which create incompatible sites in the data. Entries `D' in the table (right panel) shows the combination of left and right lineage states that lead to the ARG being recoverable (if the loop closes when the two lineages are in the corresponding states). Entry `S' indicates the start position of the two lineages directly after the recombination event.}
  \label{fig:diagram of state changes}
 \end{figure} 

Let 
\begin{equation*}
    P_3^{\rho, R}(0, n_l,n_f,r_0,r,a,b,c_1,c_2,d_1,d_2,e_1,e_2,i,j,k_1,k_2,l_1,l_2,m_1,m_2)
\end{equation*} 
be the probability that the ARG is recoverable, given that $r$ recombinations have occurred in the history, $r_0$ of which are currently open, and $n_l$ lineages remain, there are $a$ recombination loops with left lineage in State 1, $b$ in State 2, and so on, and there are no unresolved edges (denoted by the first index being 0). For the ARG to be recoverable, in each recombination loop we require that at least one of the recombinant lineages is in state 6 or 7, or both lineages in a state greater than 3, before the two recombinant lineages can coalesce (and close the recombination loop). These conditions, for each recombination loop, are necessary and sufficient: any internal branch without at least one mutation would lead to non-recoverability, as illustrated in Figure \ref{fig:missed interior branches}. Similarly, any recombination loop that closes before the recombinant lineages have reached suitable states will not generate incompatible sites, and hence the recombination will not be detectable. The recursions for this system are described in full in the Appendix, Section \ref{app_full_equation}, and the recursion solved in MATLAB to find $P_3^{\rho, R}(0,n,n,0,0,0,0,0,0,0,0,0,0,0,0,0,0,0,0,0,0)$.

\subsection{Probability that ARG topology is partially recoverable}\label{incomplete_ARG}

We also consider the probability that the ARG is \emph{partially recoverable}, with mutations present on all except (up to) $n_0$ internal edges. This is done similarly to the case for binary trees described in Section \ref{ch_trees}, and simply requires more bookkeeping; the full system of resulting equations is presented in the Appendix, Section \ref{app_full_equation}.

\subsection{One recombination}

We first consider the results when conditioning on one recombination, setting $R=1$; this is a realistic scenario when analysing sequencing data from species with a low recombination rate. We fix $\rho=0.1$, being suitably small so that the assumption of a single recombination is reasonable, noting that this is the estimated value of the recombination rate for SARS-CoV-2 genomes, and human samples of length 1kb. 

\begin{figure}[htbp!]
  \centering
  \begin{subfigure}[t]{0.5\linewidth}
    \includegraphics[trim={0 0 0 .9cm},clip,width=\linewidth]{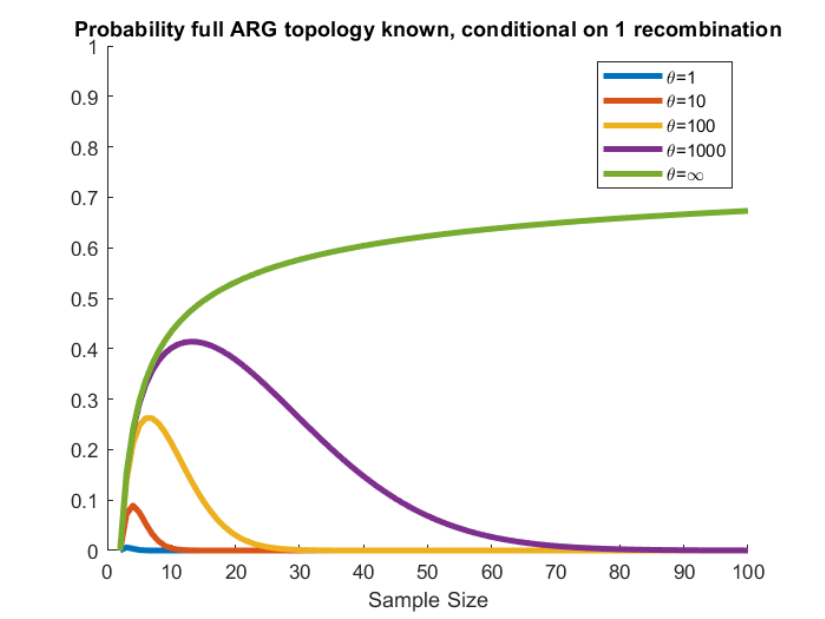}
    \caption{Breakpoint fixed at $z=0.5$, varying $n$ for various values of $\theta$ (colours).}
  \end{subfigure}
  \quad
  \begin{subfigure}[t]{0.45\linewidth}
    \includegraphics[trim={4cm 9.2cm 4cm 9.2cm}, clip, width=\linewidth]{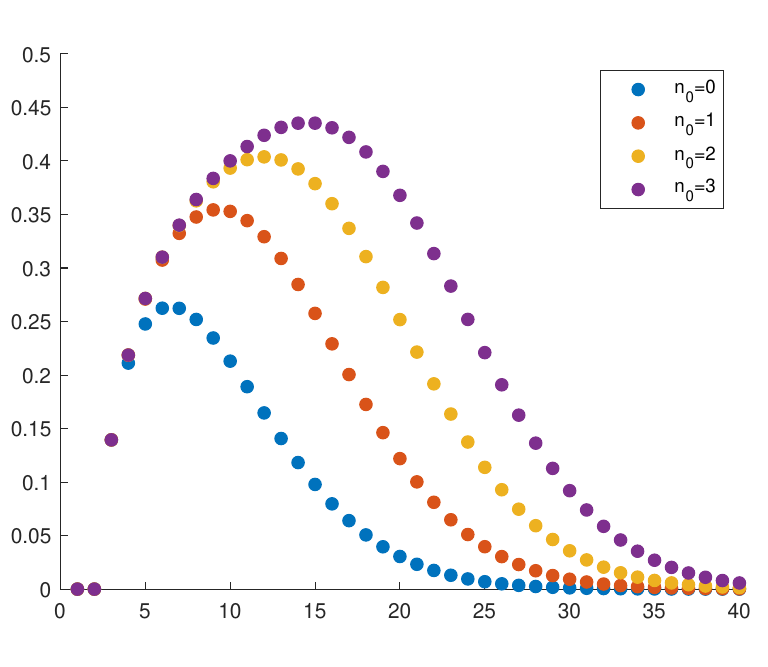}
    \caption{Fixed $\theta = 100$, varying $n$, for various numbers $n_0$ of unresolved internal edges (colours).}
  \end{subfigure}
  \caption{Probabilities of the ARG being (partially) recoverable conditional on one recombination.}
  \label{fig:full_top_with_1_recomb_graphs}
\end{figure}

Figure \ref{fig:full_top_with_1_recomb_graphs} (a) shows solutions of the recursive system for various values of the parameters. Note that these curves are not monotonic in $n$: there must be a sufficient number of coalescences above the recombination to create incompatible sites in the sample, but increasing the number of lineages makes it unlikely that a mutation occurs between each coalescence (required to make the ARG recoverable). The results demonstrate that the probability of the ARG being recoverable is very low for even moderate values of $\theta$, increasing very slowly as $\theta \to \infty$.

Figure \ref{fig:full_top_with_1_recomb_graphs}  (b) demonstrates that allowing just a small number of `missed' internal edges substantially improves the probability of recovering the rest of the ARG topology correctly. For instance, with $n=15$, the probability of recovering the ARG topology increases from around 0.1 to 0.4 if up to three unresolved internal edges are allowed. 

Solutions to the recursive system while varying the breakpoint across the genome, $z$, show that taking a breakpoint close to the centre of the genome gives slightly higher probabilities of detecting the full topology (see Figure \ref{fig:one_recomb_vary_z} in the Appendix). 

\subsection{Two recombinations}

While ARGs containing only one recombination are trivially galled, Figure \ref{fig:is_tree_galled?} shows that around 40\% of ARGs with two recombination nodes will be galled for $\rho=0.1$. The probability of an ARG being a galled tree falls substantially when conditioning on more than two recombinations, so we do not analyse this case in further detail.

\begin{figure}[htbp!]
  \centering
  \begin{subfigure}[t]{0.45\linewidth}
    \includegraphics[trim={0 0 0 .05cm}, clip, width=\linewidth]{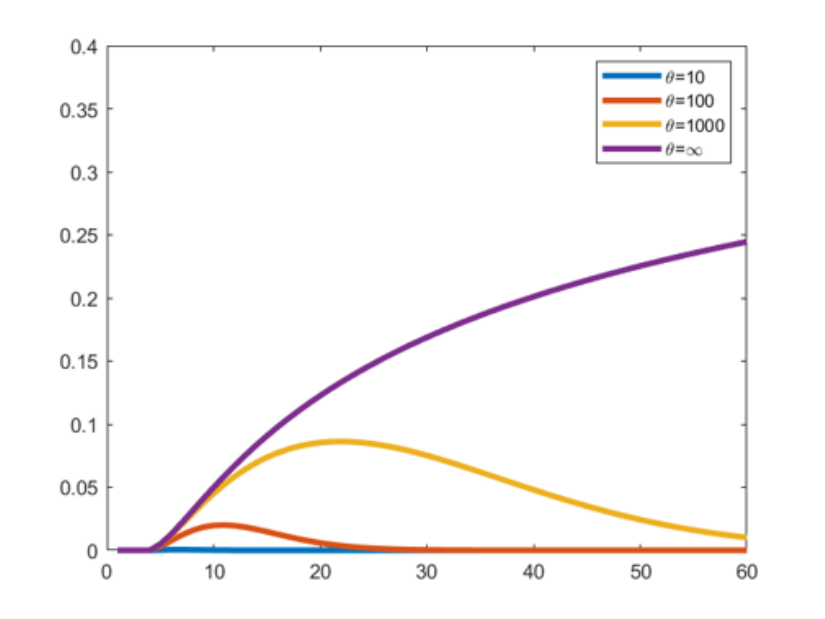}
    \caption{Breakpoint fixed at $z=0.5$. Varying $n$ for various values of $\theta$ (colours).}
  \end{subfigure}
  \quad
  \vspace{10pt}
  \begin{subfigure}[t]{0.4\linewidth}
    \includegraphics[trim={4cm 9.2cm 4cm 9.2cm},clip,width=\linewidth]{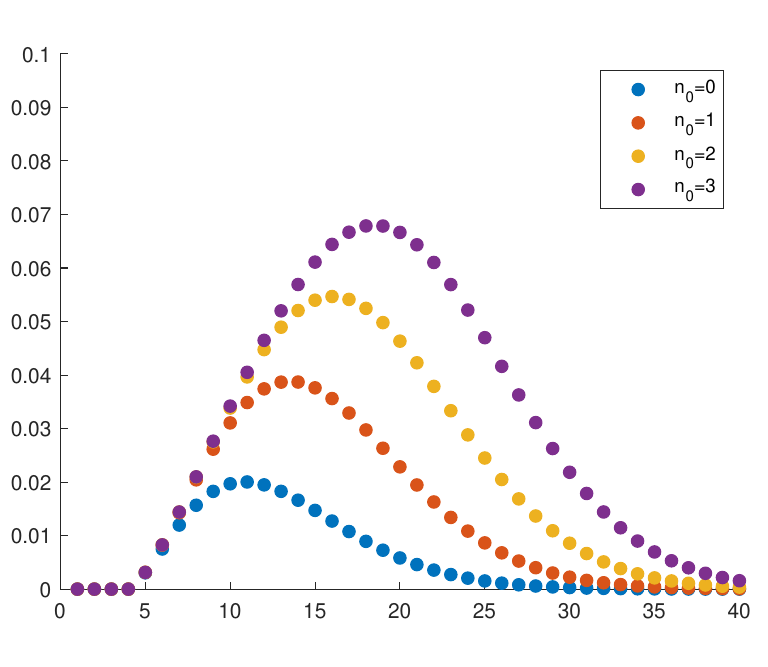}
    \caption{Fixed $\theta=100$, varying $n$, for various numbers $n_0$ of unresolved internal edges (colours). }
  \end{subfigure}
  \caption{Probabilities of the ARG being (partially) recoverable, conditioning on two galled recombinations.}
  \label{fig:full_top_with_2_recomb_graphs}
\end{figure}

Figure \ref{fig:full_top_with_2_recomb_graphs} illustrates solutions of the recursion equations when conditioning on two recombinations and the ARG being a galled tree. The probabilities of the ARG being recoverable are significantly smaller, with fewer than a quarter of ARGs being recoverable even with an infinite mutation rate, for $n = 60$. In comparison, this probability is closer to 0.7 when conditioning on only one recombination. Figure \ref{fig:full_top_with_2_recomb_graphs} (b) demonstrates again that the probability of the ARG being \emph{partially} recoverable, allowing even a small number of unresolved edges, is comparatively higher, but still very low. 

These results imply that even if an ARG reconstruction algorithm utilises all of the available information on shared mutations contained within the sequencing data, there is still likely to be significant uncertainty in resolving the location of internal edges. This probability only decreases with increasing recombination rate, and improves very slowly with increasing mutation rate. This makes it very unlikely that reconstruction programs will successfully capture the full complexity of the ARG.

The parameter values $\theta \approx 100$ and $\rho \approx 0.1$, reasonable for SARS-CoV-2, might appear to be optimal for creating genealogies that are fully recoverable from the data: low recombination rates increase the probability of seeing a small number of galled cycles, and high mutation rates make it more likely that mutations will fall on all of the necessary edges. However, our results show that at most $25\%$ of one-recombination, and $2.5\%$ of two-recombination ARGS are recoverable.

\section{Probability that gene conversion is detectable} \label{ch_geneconv}

\begin{figure}[htbp!]
  \centering
    \includegraphics[width=0.7\textwidth]{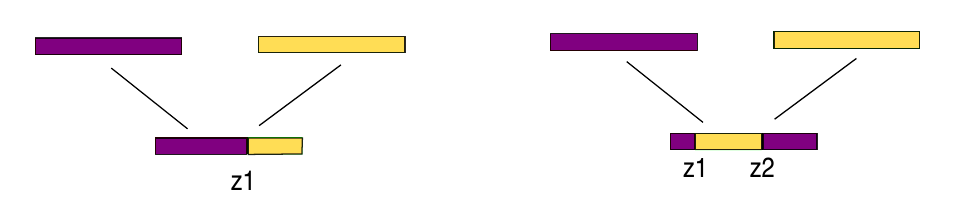}
    \caption{Crossover recombination with breakpoint at position $z_1$ (left panel), and gene conversion with conversion tract between positions $z_1$ and $z_2$ (right panel).}
   \label{fig:gene_conversion}
\end{figure}

We have so far focussed on crossover recombination events, with one fixed breakpoint: gene conversion is another important type of recombination, which is thought to commonly occur in biological settings, but which has not received as much consideration from a theoretical perspective \citep{Song06}. Figure \ref{fig:gene_conversion} illustrates the key difference between crossover recombination (left panel) and gene conversion events (right panel). Genetic material ancestral to the orange section is taken from the right parent and material ancestral to the purple section from the left. In biological samples, the conversion tract (orange) is typically small compared to the total length of the genome. 

In this section, conditioning on precisely one gene conversion event in the history of the sample (again in the \emph{big} ARG sense), we calculate the probability that this event is detectable, and is distinguishable from one crossover recombination event (\emph{without} the requirement that the ARG is recoverable).
 As we thus consider a simpler case, we can begin constructing the recursion equations immediately after (above) the gene conversion event, conditioning on precisely one such event in the genealogy. 

\begin{figure}[htbp!]
  \centering
    \includegraphics[scale=.8]{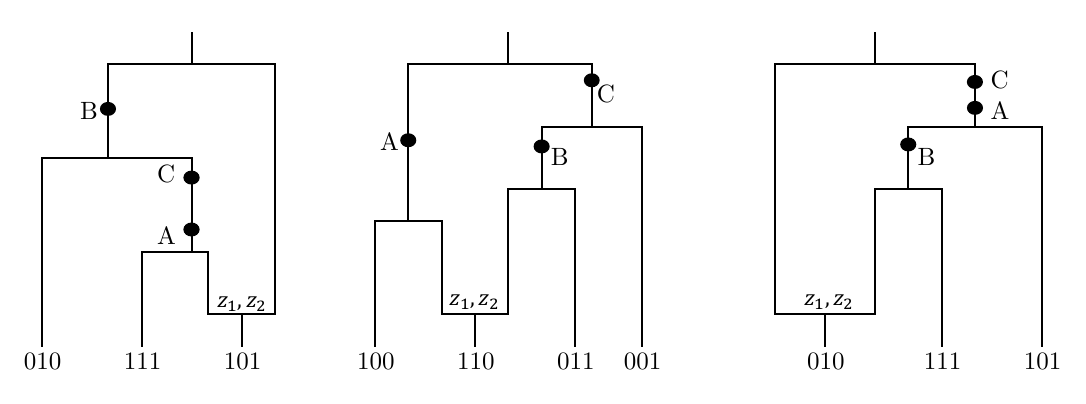}
    \caption{Conditions for a gene conversion to be detectable. Gene conversion nodes are labelled with the breakpoint positions. Each mutation is labelled by the section of the genome on which it must occur.}
   \label{fig:conversion_detection_set_ups}
\end{figure}

Let $\rho$ now be the population scaled rate of gene conversion. Similarly to the case of crossover recombination, a gene conversion is detectable if two pairs of sites spanning the two breakpoints are incompatible (in the sense of the four gamete test). Label the sections of the genome undergoing the gene conversion $[0,z_1)$, $(z_1,z_2)$, $(z_2,1]$ as $A, B, C,$ respectively. Figure \ref{fig:conversion_detection_set_ups} demonstrates the possible configurations of events inside the gene conversion loop that lead to detectability. Following similar arguments to those of \citet{Myersthesis} for the case of a single recombination, if one of the three possibilities illustrated in Figure 13 appears as a subgraph of the ARG, the gene conversion is guaranteed to be detectable. Note that there is some flexibility in the arrangement of events, as the positions of $A$ and $C$ can be interchanged, and additional coalescence events can be added to the recombination loop. Note also that the sub-graphs corresponding to sections $[0,z_2)$ and $(z_1,1]$ of the genome each must have one of the configurations given in Figure \ref{fig:configurations of mutations leading to detectable recombinations}.

\subsection{Probability gene conversion is detectable if it occurs when there are \texorpdfstring{$k-1$}{k-1} lineages in the ARG}
The recursion relations for this scenario take a similar form to those described in Section \ref{ch_s_myers}, conditioning on a single gene conversion in the genealogy, occurring at time $t$ and resulting in a transition from $k - 1 \leq n$ to $k$ lineages in the ARG. Let $i$ (resp.~$j$) denote the state of the left (resp.~right) recombinant lineage, as detailed in full in the Appendix, Tables \ref{tab:table3} and \ref{tab:table4}. At time $t' \geq t$, let $P_4^{\rho}(n_l, i, j)$ denote the probability that the gene conversion will be detectable, given there are currently $n_l$ lineages in the ARG (including the two recombinant lineages), with the left recombinant lineage $\mathcal{E}$ being in state $i$ and the right lineage $\mathcal{F}$ in state $j$. Note that unlike the case of crossover recombination, there is now a broken symmetry as two mutations on the flanking parts of the genome (purple segments in Figure \ref{fig:gene_conversion}) are needed, and only one on the conversion tract (orange segment), so the set of possible states differs for $\mathcal{E}$ and $\mathcal{F}$.

The full system of recursions is included in the Appendix, Section \ref{app_gene_conversion}. These equations are formed by considering the next state that could be reached in the ARG, and applying the law of total probability. 
As some events will not change the state of the ARG, the recursive equations can be expressed as 
\begin{align*}
    \text{(total rate of events that change the ARG)} \cdot P_4^{\rho}(n_l, i, j) &= \\
    \sum_{i',j'}\text{rate of event that results in transition between states} &(i, j) \rightarrow (i',j') \cdot P_4^{\rho}(n_l', i', j').
\end{align*}
Events which change the state of the ARG include coalescences and mutations (as the position of the gene conversion event is separately conditioned upon). Each equation takes the general form
\begin{align*}
    \Bigg( \binom{n_l}{2} + g(\theta) + \rho \frac{n_l}{2} \Bigg) P_4^{\rho}(n_l, i, j) = &\binom{n_l-1}{2} P_4^{\rho}(n_l - 1, i, j) \\
    &+ (n_l-2) P_4^{\rho}(n_l-1, i', j') \\
    &+  \sum_{i',j'} g_{i',j'}(\theta) P_4^{\rho}(n_l, i', j'),
\end{align*}
where $g_{i,j}(\theta)$ are linear functions of $\theta$ which are different for each pair of states $(i,j)$, and $g_{i,j}(\theta)=\sum_{i',j'} g_{i',j'}(\theta)$. 
The recursive system is solved to find $P_4^{\rho}(k, 0, 0)$.

\subsection{Probability gene conversion is detectable}
Summing over $k$ removes the conditioning on when the gene conversion event occurs, to give the desired probability $\mathcal{P}^{n, \rho}$, that a detectable gene conversion events occurs, conditional on precisely one such event in the history. We have
\begin{align} \label{gc_equation}
    \mathcal{P}^{n, \rho}&= \frac{\sum_{k=2}^n  \left(\prod_{l=k}^n \frac{l-1}{l-1+\rho}\right) \frac{\rho}{k-2+\rho} \cdot P_4^{\rho}(k, 0, 0)} {\sum_{k=2}^n  \left(\prod_{l=k}^n \frac{l-1}{l-1+\rho}\right) \frac{\rho}{k-2+\rho} \left(\prod_{l=2}^{k} \frac{l-1}{l-1+\rho}\right)} \\ \nonumber
    &=\frac{\sum_{k=2}^n  \left(\prod_{l=2}^{k-1} \frac{l-1}{l-1+\rho}\right) \frac{\rho}{k-2+\rho} \cdot P_4^{\rho}(k, 0, 0)} {\sum_{k=2}^n \frac{\rho}{k-2+\rho}  \cdot \frac{k-1}{k-1+\rho}}.
\end{align}
This is constructed as follows. Note that $P_4^{\rho}(k, 0, 0)$ describes the state of an ARG directly after the gene conversion (looking backwards in time), which took the number of lineages from $k-1$ to  $k$. Therefore, starting with a sample of size $n$, there must have been $n-k$ coalescence events, followed by the gene conversion, followed by  the a sequence of state changes that allow the gene conversion to be detectable. These events, respectively, have probabilities
\begin{equation*}
    \prod_{l=k}^n \frac{l-1}{l-1+\rho}, \;\;\;  \frac{\rho}{k-2+\rho}, \;\;\; P_4^{\rho}(k, 0, 0).
\end{equation*}
Putting these together gives the numerator of \eqref{gc_equation}. The denominator is constructed in a similar way, requiring $n-k$ coalescent events followed by the gene conversion. After the gene conversion event, only a further $k-1$ coalescences are required, with probability 
\begin{equation*}
    \prod_{l=2}^k \frac{l-1}{l-1+\rho}.
\end{equation*}

\begin{figure}[htbp!]
  \centering
  \begin{subfigure}[t]{0.45\linewidth}
    \includegraphics[trim={4cm 9.2cm 4cm 9.2cm}, clip, width=\linewidth]{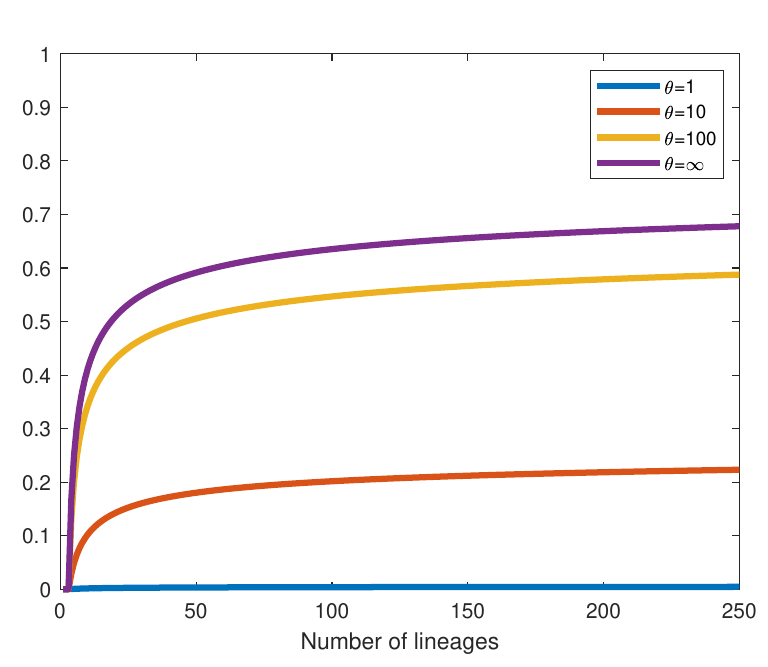}
    \caption{Varying $n$ for several fixed values of $\theta$ (colours), with breakpoints at $z=0.33, 0.67$}
  \end{subfigure}
  \quad
  \begin{subfigure}[t]{0.45\linewidth}
    \includegraphics[trim={4cm 9.2cm 4cm 9.2cm}, clip, width=\linewidth]{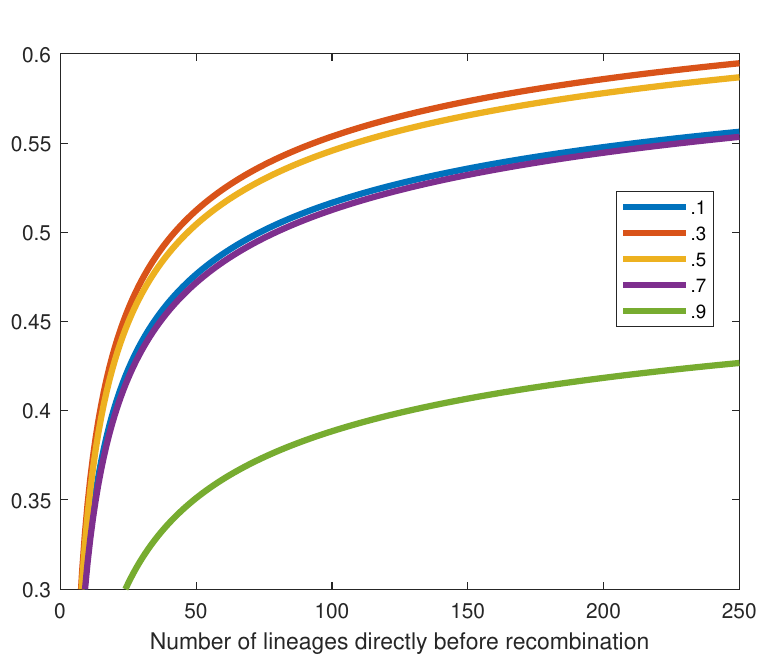}
    \caption{$\theta=100$, varying the length of the converted section. Note $y$-axis scale is truncated. Mutation rate is  uniform over the genome, the conversion tract is centred about 0.5.}
  \end{subfigure}
  \caption{Probability of gene conversion being detectable.}
  \label{fig:conversion_detection_graph}
\end{figure}

Figure \ref{fig:conversion_detection_graph} (a) shows that detection probabilities for gene conversion events behave very similarly to those for detecting a single recombination \citep{Myersthesis}, though are consistently slightly lower, as the gene conversion requires more mutation events to be detectable. Note that the asymptotic probability as $\theta \xrightarrow{} \infty$ tends to the probability that a single recombination changes the ARG topology. In Figure \ref{fig:conversion_detection_graph} (b), the length of the conversion tract is varied; for scenarios where either the conversion tract, or its complement, is particularly short, the probability of detection decreases, as there is a lower probability of a mutation falling on the shorter section. As the mutation rate is assumed to be uniform across the genome, a conversion length of $1/3$ gives the highest probabilities of detection. 

\section{Discussion} \label{ch_discussion}

In this article, we have calculated the probability of recovering the tree or ARG topology under the coalescent with recombination, when the ARG topology is constrained to be in the shape of a galled tree. Galled trees have several attractive combinatorial and algorithmic properties that do not hold for general ARGs. We have explicitly calculated the probability of an ARG being a galled tree, shedding light on how applicable these results might be in the analysis of real data. Our results indicate that genealogies in the form of galled trees are reasonably likely to be seen for $\rho < 1$ with moderate sample size.

Our results can also shed light on some theoretical properties of genealogical reconstruction algorithms. While some recently developed methods can handle impressive quantities of sequencing data, they are based on heuristic methods, making it difficult to obtain theoretical insights into their performance. In particular, while tsinfer \citep{Kelleher19} retains polytomies (i.e.\ nodes with more than two child lineages) where the order of coalescence events cannot be resolved unambiguously, many other algorithms resolve the order of events randomly. Our results give a sense of how many such polytomies might be present in the history of a dataset, and how likely recombination events are to be detectable for a given value of evolutionary parameters. This provides an upper bound on how well genealogies can be reconstructed, even if the algorithm utilises all of the available sequencing data to the fullest extent. 

In the absence of recombination, our results demonstrate that allowing a small number of unresolved internal edges can greatly improve the probability of reconstructing the rest of the tree correctly. This suggests that, for certain values of the parameters, there are likely to be a relatively small number of edges in the genealogy which are not supported by mutations and could be placed at many plausible positions. 

For large sample sizes, the probability of recovering the tree or ARG topology with a high level of certainty is minuscule, for reasonable values of the mutation rate (such as those estimated for SARS-CoV-2). This strengthens the case for using Bayesian methods to integrate over the uncertainty of branch placements, or utilising additional data to resolve ambiguous event ordering. For instance, \citet{ramazotti} analysed variant frequencies using SARS-CoV-2 intra-host sequencing data, in order to resolve the ordering of transmission events in genealogies built using consensus sequences (i.e.\ at the level of one sequence per infected host).

For tractability, our analysis has focused on the particular case where the ARG topology is that of a galled tree, under the coalescent with recombination. A natural extension of this work would be to consider general ARGs and other models, with more complex scenarios that might include multiple loci or non-constant population size. 

\section*{Acknowledgements}
We thank Paul Jenkins for useful comments, and two anonymous reviewers for valuable suggestions. This work was supported by the EPSRC and MRC OxWaSP Centre for Doctoral Training (EPSRC grant EP/L016710/1), and by the Alan Turing Institute (EPSRC grant EP/N510129/1).

\bibliographystyle{statsy}
\bibliography{bibliography.bib}

\begin{thebibliography}{33}
\providecommand{\natexlab}[1]{#1}
\providecommand{\url}[1]{\texttt{#1}}
\providecommand{\urlprefix}{URL }
\expandafter\ifx\csname urlstyle\endcsname\relax
  \providecommand{\doi}[1]{doi:\discretionary{}{}{}#1}\else
  \providecommand{\doi}{doi:\discretionary{}{}{}\begingroup
  \urlstyle{rm}\Url}\fi

\bibitem[Buneman, 1971]{buneman}
Buneman, P. (1971).
\newblock The recovery of trees from measures of dissimilarity.
\newblock In F.~R. Hodson, D.~G. Kendall and P.~Tautu, editors,
  \emph{Mathematics in the Archeological and Historical Sciences}, pp.
  387--395. Edinburgh University Press, Edinburgh.

\bibitem[Chan et~al., 2012]{Chan12}
Chan, A.~H., Jenkins, P.~A. and Song, Y.~S. (2012).
\newblock Genome-wide fine-scale recombination rate variation in {D}rosophila
  melanogaster.
\newblock \emph{PLoS Genetics}, \textbf{8}(12), e1003090.

\bibitem[Duchene et~al., 2020]{Duchene20}
Duchene, S., Featherstone, L., Haritopoulou-Sinanidou, M., Rambaut, A., Lemey,
  P. and Baele, G. (2020).
\newblock Temporal signal and the phylodynamic threshold of {SARS-CoV-2}.
\newblock \emph{Virus Evolution}, \textbf{6}(2).

\bibitem[Ethier and Griffiths, 1990]{ethier1990two}
Ethier, S. and Griffiths, R. (1990).
\newblock On the two-locus sampling distribution.
\newblock \emph{Journal of Mathematical Biology}, \textbf{29}(2), 131--159.

\bibitem[Fu and Li, 1993]{Fu693}
Fu, Y.-X. and Li, W.-H. (1993).
\newblock Statistical tests of neutrality of mutations.
\newblock \emph{Genetics}, \textbf{133}(3), 693--709.

\bibitem[Griffiths and Marjoram, 1997]{griffithsmarjoram}
Griffiths, R.~C. and Marjoram, P. (1997).
\newblock An ancestral recombination graph.
\newblock In P.~Donnelly and S.~Tavare, editors, \emph{Progress in population
  genetics and human evolution}, pp. 257--270. Springer, New York.

\bibitem[Gusfield, 2014]{Gusfield08}
Gusfield, D. (2014).
\newblock \emph{ReCombinatorics: The algorithmics of ancestral recombination
  graphs and explicit phylogenetic networks}.
\newblock MIT press, Cambridge, Massachusetts.

\bibitem[Gusfield et~al., 2004]{gusfield04}
Gusfield, D., Eddhu, S. and Langley, C. (2004).
\newblock Optimal, efficient reconstruction of phylogenetic networks with
  constrained recombination.
\newblock \emph{Journal of Bioinformatics and Computational Biology},
  \textbf{2}(01), 173--213.

\bibitem[Heath et~al., 2008]{heath2008taxon}
Heath, T.~A., Hedtke, S.~M. and Hillis, D.~M. (2008).
\newblock Taxon sampling and the accuracy of phylogenetic analyses.
\newblock \emph{Journal of Systematics and Evolution}, \textbf{46}(3), 239.

\bibitem[Hillis, 1998]{hillis1998taxonomic}
Hillis, D.~M. (1998).
\newblock Taxonomic sampling, phylogenetic accuracy, and investigator bias.
\newblock \emph{Systematic Biology}, \textbf{47}(1), 3--8.

\bibitem[Hillis et~al., 1994]{hillis1994hobgoblin}
Hillis, D.~M., Huelsenbeck, J.~P. and Swofford, D.~L. (1994).
\newblock Hobgoblin of phylogenetics?
\newblock \emph{Nature}, \textbf{369}(6479), 363--364.

\bibitem[Hobolth and Wiuf, 2009]{Hobolth2009TheGS}
Hobolth, A. and Wiuf, C. (2009).
\newblock The genealogy, site frequency spectrum and ages of two nested mutant
  alleles.
\newblock \emph{Theoretical Population Biology}, \textbf{75 4}, 260--5.

\bibitem[Hudson and Kaplan, 1985]{Hudson85}
Hudson, R.~R. and Kaplan, N.~L. (1985).
\newblock Statistical properties of the number of recombination events in the
  history of a sample of {DNA} sequences.
\newblock \emph{Genetics}, \textbf{111}(1), 147--164.

\bibitem[Jenkins et~al., 2014]{jenkins14}
Jenkins, P.~A., Mueller, J.~W. and Song, Y.~S. (2014).
\newblock General triallelic frequency spectrum under demographic models with
  variable population size.
\newblock \emph{Genetics}, \textbf{196}, 295–311.

\bibitem[Jenkins and Song, 2009]{jenkins2009closed}
Jenkins, P.~A. and Song, Y.~S. (2009).
\newblock Closed-form two-locus sampling distributions: accuracy and
  universality.
\newblock \emph{Genetics}, \textbf{183}(3), 1087--1103.

\bibitem[Jenkins and Song, 2010]{jenkins2010asymptotic}
Jenkins, P.~A. and Song, Y.~S. (2010).
\newblock An asymptotic sampling formula for the coalescent with recombination.
\newblock \emph{The Annals of Applied Probability}, \textbf{20}(3), 1005.

\bibitem[Jenkins and Song, 2011]{jenkins11}
Jenkins, P.~A. and Song, Y.~S. (2011).
\newblock The effect of recurrent mutation on the frequency spectrum of a
  segregating site and the age of an allele.
\newblock \emph{Theoretical Population Biology}, \textbf{80}(2), 158--173.

\bibitem[Kelleher et~al., 2019]{Kelleher19}
Kelleher, J., Wong, Y., Wohns, A.~W., Fadil, C., Albers, P.~K. and McVean, G.
  (2019).
\newblock Inferring whole-genome histories in large population datasets.
\newblock \emph{Nature Genetics}, \textbf{51}(9), 1330--1338.

\bibitem[Kim, 1998]{kim1998large}
Kim, J. (1998).
\newblock Large-scale phylogenies and measuring the performance of phylogenetic
  estimators.
\newblock \emph{Systematic Biology}, \textbf{47}(1), 43--60.

\bibitem[Li et~al., 2020]{li_gentime}
Li, Q., Guan, X., Wu, P., Wang, X., Zhou, L., Tong, Y., Ren, R., Leung, K.~S.,
  Lau, E.~H., Wong, J.~Y. et~al. (2020).
\newblock Early transmission dynamics in {W}uhan, {C}hina, of novel
  coronavirus--infected pneumonia.
\newblock \emph{New England {J}ournal of {M}edicine}, \textbf{382}, 1199--1207.

\bibitem[M\"uller et~al., 2021]{Muller21}
M\"uller, N.~F., Kistler, K.~E. and Bedford, T. (2021).
\newblock Recombination patterns in coronaviruses.
\newblock \emph{bioRxiv}.
\newblock \doi{10.1101/2021.04.28.441806}.

\bibitem[Myers, 2003]{Myersthesis}
Myers, S. (2003).
\newblock \emph{The detection of recombination events using {DNA} sequence
  data}.
\newblock Ph.D. thesis, University of Oxford, Department of Statistics.

\bibitem[Pollock et~al., 2002]{pollock2002increased}
Pollock, D.~D., Zwickl, D.~J., McGuire, J.~A. and Hillis, D.~M. (2002).
\newblock Increased taxon sampling is advantageous for phylogenetic inference.
\newblock \emph{Systematic biology}, \textbf{51}(4), 664.

\bibitem[Ramazzotti et~al., 2021]{ramazotti}
Ramazzotti, D., Angaroni, F., Maspero, D., Gambacorti-Passerini, C.,
  Antoniotti, M., Graudenzi, A. and Piazza, R. (2021).
\newblock Verso: a comprehensive framework for the inference of robust
  phylogenies and the quantification of intra-host genomic diversity of viral
  samples.
\newblock \emph{Patterns}, \textbf{2}(3), 100212.

\bibitem[Rasmussen et~al., 2014]{argweaver}
Rasmussen, M.~D., Hubisz, M.~J., Gronau, I. and Siepel, A. (2014).
\newblock Genome-wide inference of ancestral recombination graphs.
\newblock \emph{PLoS Genetics}, \textbf{10}(5), e1004342.

\bibitem[Sargsyan, 2006]{sargsyan2006analytical}
Sargsyan, O. (2006).
\newblock \emph{Analytical and simulation results for the general coalescent}.
\newblock Ph.D. thesis, University of Southern California.

\bibitem[Semple and Steel, 2003]{semple2003phylogenetics}
Semple, C. and Steel, M. (2003).
\newblock \emph{Phylogenetics}.
\newblock Oxford University Press, Oxford.

\bibitem[Song, 2006]{songgalledtrees}
Song, Y.~S. (2006).
\newblock A concise necessary and sufficient condition for the existence of a
  galled-tree.
\newblock \emph{IEEE/ACM Transactions on Computational Biology and
  Bioinformatics}, \textbf{3}(2), 186--191.

\bibitem[Song et~al., 2008]{Song06}
Song, Y.~S., Ding, Z., Gusfield, D., Langley, C.~H. and Wu, Y. (2008).
\newblock Algorithms to distinguish the role of gene-conversion from
  single-crossover recombination in the derivation of {SNP} sequences in
  populations.
\newblock \emph{Journal of Computational Biology}, \textbf{14}(10), 1273--86.

\bibitem[Song and Hein, 2005]{Hein05}
Song, Y.~S. and Hein, J. (2005).
\newblock Constructing minimal ancestral recombination graphs.
\newblock \emph{Journal of Computational Molecular Cell Biology},
  \textbf{12}(2), 147--169.

\bibitem[Speidel et~al., 2019]{relate}
Speidel, L., Forest, M., Shi, S. and Myers, S. (2019).
\newblock A method for genome-wide genealogy estimation for thousands of
  samples.
\newblock \emph{Nature Genetics}, \textbf{51}(9), 1321--1329.

\bibitem[Wang et~al., 2001]{wang01}
Wang, L., Zhang, K. and Zhang, L. (2001).
\newblock Perfect phylogenetic networks with recombination.
\newblock \emph{Journal of Computational Biology}, \textbf{8}(1), 69--78.

\bibitem[Wiuf and Donnelly, 1999]{wiuf99}
Wiuf, C. and Donnelly, P. (1999).
\newblock Conditional genealogies and the age of a neutral mutant.
\newblock \emph{Theoretical Population Biology}, \textbf{56}(2), 183--201.

\end{thebibliography}

\newpage
\appendix
\addappheadtotoc

\newgeometry{right=15mm, top=25mm, left=15mm, bottom=25mm}
\pagenumbering{roman}
\setcounter{page}{1}
\section{Supplementary Material}

\subsection{Model for full recoverability of the ARG} \label{app_full_equation}
\vspace{.7cm}
\small

This section presents the recursion equations for the probability of the ARG being a galled tree and recoverable, conditional on $R$ recombinations. If a uniform mutation rate along the genome is assumed, with total mutation rate $\theta /2$ and a breakpoint at position $z \in [0,1]$; then type $A$ mutations occur at rate $\theta_A/2 := z \cdot \theta/2$ and type $B$ with rate $\theta_B/2 := (1-z) \cdot \theta/2$. It should be noted that the specific placement of the mutation within each locus does not matter.

For the sake of clarity, the subscript indices are contracted so that only those changing at each step are shown. For instance, 
\begin{align*}
\Tilde{P}_3^{\rho, R}(n_0,n_l-1,n_f-1,r_0,r, i-1,j+1)= P_3^{\rho, R}(n_0,n_l-1,&n_f-1,r_0,r, a,b,c_1,c_2,d_1,d_2,e_1,e_2,\\ &i-1,j+1,k_1,k_2,l_1,l_2,m_1,m_2),
\end{align*}
and $\Tilde{P}_3^{\rho, R}(n_0,n_l,n_f+1,r_0,r)$ indicates that none of the recombinant states have changed.

The index $r$ tracks the number of open recombination loops, so there is a restriction 
\[
r = a+b+c_1+c_2+d_1+d_2+e_1+e_2 = i+j+k_1+k_2+l_1+l_2+m_1+m_2,
\]
with $2r_0 \leq n_l \leq n + r_0$. For any values of the indices that do not meet this condition, $\Tilde{P}_3^{\rho, R}(\cdot) = 0$.

For clarity, the main equation is broken up into  several parts. As stated in the main text, the equations take the form 
\begin{multline*}
\text{(total rate of moves that  change ARG state)} \cdot \Tilde{P}_3^{\rho, R}(n_0,n_l-1,n_f-1,r_0,r) = \\
\sum \textrm{[rate of event that results in transition from  states $(i,j)$}] \cdot \Tilde{P}_3^{\rho, R}(\text{resultant state}).
\end{multline*}
The total rate of moves that change the state of the ARG is
\begin{multline*}
    Rate=\Bigg(\binom{n_l}{2}+ \frac{\rho n_l}{2} + \frac{\theta}{2}(n_l-n_f-2r_0 + d_2 + l_2)\\ 
    + \frac{\theta_A}{2}(b+d_1+e_2+j+k_2+l_1)+\frac{\theta_B}{2}(b+c_2+d_1+j+l_1+m_2)\Bigg).
\end{multline*}
Moves that change the state of the ARG are as follows:
\begin{enumerate}
    \item Coalescence of non-recombinant lineages, with rate
\begin{multline*}
    Coal_{NR}=\binom{n_f}{2}\Tilde{P}_3^{\rho, R}(n_0,n_l-1,n_f-2,r_0,r) + {n_f(n-2r_0-n_f)}\Tilde{P}_3^{\rho, R}(n_0-1,n_l-1,n_f-1,r_0,r) \\
    + \binom{n-2r_0-n_f}{2}\Tilde{P}_3^{\rho, R}(n_0-2,n_l-1,n_f,r_0,r).
\end{multline*}

\item The first mutation of a non-recombinant lineage since its last coalescence,
\begin{equation*}
    Mut_{NR}=(n_l-n_f-2r_0)\frac{\theta}{2} \Tilde{P}_3^{\rho, R}(n_0,n_l,n_f+1,r_0,r).
\end{equation*}

\item A coalescence of one recombinant lineage, and one non-recombinant (taking care to distinguish whether the non-recombinant lineage has had a mutation since its last coalescence),
\begin{multline*}
    Coal_{R}= i \cdot n_f \cdot \Tilde{P}_3^{\rho, R}(n_0,n_l-1,n_f-1,r_0,r,i-1,j+1) \\+ i \cdot (n-2r_0-n_f) \cdot \Tilde{P}_3^{\rho, R}(n_0-1,n_l-1,n_f,r_0,r,i-1,j+1) + a \cdot n_f \cdot \Tilde{P}_3^{\rho, R}(n_0,n_l-1,n_f-1,r_0,r,a-1,b+1) \\+ a \cdot (n-2r_0-n_f) \cdot \Tilde{P}_3^{\rho, R}(n_0-1,n_l-1,n_f,r_0,r,a-1,b+1) + k_1 \cdot n_f \cdot \Tilde{P}_3^{\rho, R}(n_0,n_l-1,n_f-1,r_0,r, k_1-1,l_1+1) \\+ k_1 \cdot (n-2r_0-n_f) \cdot \Tilde{P}_3^{\rho, R}(n_0-1,n_l-1,n_f,r_0,r, k_1-1,l_1+1) + c_1 \cdot n_f \cdot \Tilde{P}_3^{\rho, R}(n_0,n_l-1,n_f-1,r_0,r, c_1-1,d_1+1) \\+ c_1 \cdot (n-2r_0-n_f) \cdot \Tilde{P}_3^{\rho, R}(n_0-1,n_l-1,n_f,r_0,r, c_1-1,d_1+1) 
    + k_2 \cdot n_f \cdot \Tilde{P}_3^{\rho, R}(n_0,n_l-1,n_f-1,r_0,r, k_1+1,k_2-1)\\ + k_2 \cdot (n-2r_0-n_f) \cdot \Tilde{P}_3^{\rho, R}(n_0-1,n_l-1,n_f,r_0,r, k_1+1,k_2-1) + c_2 \cdot n_f \cdot \Tilde{P}_3^{\rho, R}(n_0,n_l-1,n_f-1,r_0,r, b+1,c_2-1) \\+ c_2 \cdot (n-2r_0-n_f) \cdot \Tilde{P}_3^{\rho, R}(n_0-1,n_l-1,n_f,r_0,r, b+1,c_2-1)
    + m_2 \cdot n_f \cdot \Tilde{P}_3^{\rho, R}(n_0,n_l-1,n_f-1,r_0,r, l_1+1,m_2-1) \\+ m_2 \cdot (n-2r_0-n_f) \cdot \Tilde{P}_3^{\rho, R}(n_0-1,n_l-1,n_f,r_0,r, l_1+1,m_2-1) + e_2 \cdot n_f \cdot \Tilde{P}_3^{\rho, R}(n_0,n_l-1,n_f-1,r_0,r, d_1+1,e_2-1) \\+ e_2 \cdot (n-2r_0-n_f) \cdot \Tilde{P}_3^{\rho, R}(n_0-1,n_l-1,n_f,r_0,r, d_1+1,e_2-1) 
    + m_1 \cdot n_f \cdot \Tilde{P}_3^{\rho, R}(n_0,n_l-1,n_f-1,r_0,r, l_1+1,m_1-1) \\+ m \cdot (n-2r_0-n_f) \cdot \Tilde{P}_3^{\rho, R}(n_0-1,n_l-1,n_f,r_0,r, l_1+1,m_1-1) + e_1 \cdot n_f \cdot \Tilde{P}_3^{\rho, R}(n_0,n_l-1,n_f-1,r_0,r, d_1+1,e_1-1) \\+ e_1 \cdot (n-2r_0-n_f) \cdot \Tilde{P}_3^{\rho, R}(n_0-1,n_l-1,n_f,r_0,r, d_1+1,e_1-1)
    + j \cdot n_f \cdot \Tilde{P}_3^{\rho, R}(n_0-1,n_l-1,n_f-1,r_0,r)\\ + j \cdot (n-2r_0-n_f) \cdot \Tilde{P}_3^{\rho, R}(n_0-2,n_l-1,n_f,r_0,r) + b \cdot n_f \cdot \Tilde{P}_3^{\rho, R}(n_0-1,n_l-1,n_f-1,r_0,r) \\+ b \cdot (n-2r_0-n_f) \cdot \Tilde{P}_3^{\rho, R}(n_0-2,n_l-1,n_f,r_0,r)
    + l_1 \cdot n_f \cdot \Tilde{P}_3^{\rho, R}(n_0-1,n_l-1,n_f-1,r_0,r) \\+ l_1 \cdot (n-2r_0-n_f) \cdot \Tilde{P}_3^{\rho, R}(n_0-2,n_l-1,n_f,r_0,r) + l_2 \cdot n_f \cdot \Tilde{P}_3^{\rho, R}(n_0-1,n_l-1,n_f-1,r_0,r)\\ + l_2 \cdot (n-2r_0-n_f) \cdot \Tilde{P}_3^{\rho, R}(n_0-2,n_l-1,n_f,r_0,r)
    + d_1 \cdot n_f \cdot \Tilde{P}_3^{\rho, R}(n_0-1,n_l-1,n_f-1,r_0,r) \\+ d_1 \cdot (n-2r_0-n_f) \cdot \Tilde{P}_3^{\rho, R}(n_0-2,n_l-1,n_f,r_0,r) + d_2 \cdot n_f \cdot \Tilde{P}_3^{\rho, R}(n_0-1,n_l-1,n_f-1,r_0,r) \\+ d_2 \cdot (n-2r_0-n_f) \cdot \Tilde{P}_3^{\rho, R}(n_0-2,n_l-1,n_f,r_0,r).
\end{multline*}

\item A mutation event that changes the state of a recombinant lineage,
\begin{multline*}
    Mut_{R}=j \cdot \frac{\theta_A}{2} \Tilde{P}_3^{\rho, R}(n_0,n_l,n_f,r_0,r, j-1,k_1+1) +b \cdot \frac{\theta_B}{2} \Tilde{P}_3^{\rho, R}(n_0,n_l,n_f,r_0,r, b-1,c_1+1) \\+k_2 \cdot \frac{\theta_A}{2} \Tilde{P}_3^{\rho, R}(n_0,n_l,n_f,r_0,r, k_1+1,k_2-1) +c_2 \cdot \frac{\theta_B}{2}\Tilde{P}_3^{\rho, R}(n_0,n_l,n_f,r_0,r, c_1+1,c_2-1) \\+j \cdot \frac{\theta_B}{2} \Tilde{P}_3^{\rho, R}(n_0,n_l,n_f,r_0,r, j-1,k_2+1) +b \cdot \frac{\theta_A}{2} \Tilde{P}_3^{\rho, R}(n_0,n_l,n_f,r_0,r, b-1,c_2+1)  \\+l_1 \cdot \frac{\theta_B}{2} \Tilde{P}_3^{\rho, R}(n_0,n_l,n_f,r_0,r, l_1-1,m_1+1) +d_1 \cdot \frac{\theta_A}{2} \Tilde{P}_3^{\rho, R}(n_0,n_l,n_f,r_0,r, d_1-1,e_1+1) \\+m_2 \cdot \frac{\theta_B}{2} \Tilde{P}_3^{\rho, R}(n_0,n_l,n_f,r_0,r, m_1+1,m_2-1) +e_2 \cdot \frac{\theta_A}{2} \Tilde{P}_3^{\rho, R}(n_0,n_l,n_f,r_0,r, e_1+1,e_2-1)\\ +l_1 \cdot \frac{\theta_A}{2} \Tilde{P}_3^{\rho, R}(n_0,n_l,n_f,r_0,r, l_1-1,m_2+1)+d_1 \cdot \frac{\theta_B}{2} \Tilde{P}_3^{\rho, R}(n_0,n_l,n_f,r_0,r, d_1-1,e_2+1)
    \\+l_2 \cdot \frac{\theta}{2} \Tilde{P}_3^{\rho, R}(n_0,n_l,n_f,r_0,r, l_1-2,m_1+1)+d_2 \cdot \frac{\theta}{2} \Tilde{P}_3^{\rho, R}(n_0,n_l,n_f,r_0,r, d_2-1,e_1+1).
\end{multline*}

\item The coalescence of two recombinant lineages, which for a galled tree must be the result of an open recombination loop closing. This requires a factor of $1/r$ in the probabilities, as each left recombinant lineage must choose to coalesce with its partner out of the $r$ possible right recombinant lineages available,
\begin{multline*}
    Coal_{RR}=\frac{l_2}{r} \cdot \Big(a \cdot \Tilde{P}_3^{\rho, R}(n_0-1,n_l-1,n_f,r_0-1,r, a-1,l_2-1) + a \cdot \Tilde{P}_3^{\rho, R}(n_0-2,n_l-1,n_f,r_0-1,r, b-1,l_2-1) \\+c_1 \cdot \Tilde{P}_3^{\rho, R}(n_0-1,n_l-1,n_f,r_0-1,r, c_1-1,l_2-1) + c_2 \cdot \Tilde{P}_3^{\rho, R}(n_0-1,n_l-1,n_f,r_0-1,r, c_2-1,l_2-1)\\
    +d_1 \cdot \Tilde{P}_3^{\rho, R}(n_0-2,n_l-1,n_f,r_0-1,r, d_1-1,l_2-1) + d_2 \cdot \Tilde{P}_3^{\rho, R}(n_0-2,n_l-1,n_f,r_0-1,r, d_2-1,l_2-1) \\+ e_2 \cdot \Tilde{P}_3^{\rho, R}(n_0-1,n_l-1,n_f,r_0-1,r, e_2-1,l_2-1) +e_1 \cdot \Tilde{P}_3^{\rho, R}(n_0-1,n_l-1,n_f,r_0-1,r, e_1-1,l_2-1)\Big)\\
    +\frac{m_1}{r} \cdot \Big(a \cdot \Tilde{P}_3^{\rho, R}(n_0,n_l-1,n_f,r_0-1,r, a-1,m_1-1) + b \cdot \Tilde{P}_3^{\rho, R}(n_0-1,n_l-1,n_f,r_0-1,r, b-1,m_1-1) \\+c_1 \cdot \Tilde{P}_3^{\rho, R}(n_0,n_l-1,n_f,r_0-1,r, c_1-1,m_1-1) + c_2 \cdot \Tilde{P}_3^{\rho, R}(n_0,n_l-1,n_f,r_0-1,r, c_2-1,m_1-1)\\
    +d_1 \cdot \Tilde{P}_3^{\rho, R}(n_0-1,n_l-1,n_f,r_0-1,r, d_1-1,m_1-1) + d_2 \cdot \Tilde{P}_3^{\rho, R}(n_0-1,n_l-1,n_f,r_0-1,r, d_2-1,m_1-1) \\+ e_2 \cdot \Tilde{P}_3^{\rho, R}(n_0,n_l-1,n_f,r_0-1,r, e_2-1,m_1-1) +e_1 \cdot \Tilde{P}_3^{\rho, R}(n_0,n_l-1,n_f,r_0-1,r, e_1-1,m_1-1)\Big)\\
    +\frac{e_1}{r} \cdot \Big(i \cdot \Tilde{P}_3^{\rho, R}(n_0,n_l-1,n_f,r_0-1,r, e_1-1,i-1) + j \cdot \Tilde{P}_3^{\rho, R}(n_0-1,n_l-1,n_f,r_0-1,r, e_1-1,j-1) \\ +k_1 \cdot \Tilde{P}_3^{\rho, R}(n_0,n_l-1,n_f,r_0-1,r, e_1-1,k_1-1) + k_2 \cdot \Tilde{P}_3^{\rho, R}(n_0,n_l-1,n_f,r_0-1,r, e_1-1,k_2-1) \\
    + m_2 \cdot \Tilde{P}_3^{\rho, R}(n_0,n_l-1,n_f,r_0-1,r, e_1-1,m_2-1) + l_1 \cdot \Tilde{P}_3^{\rho, R}(n_0-1,n_l-1,n_f,r_0-1,r, e_1-1,l_1-1)\Big)\\+\frac{e_2}{r} \cdot \Big(i \cdot \Tilde{P}_3^{\rho, R}(n_0-1,n_l-1,n_f,r_0-1,r, d_2-1,i-1) + j \cdot \Tilde{P}_3^{\rho, R}(n_0-2,n_l-1,n_f,r_0-1,r, d_2-1,j-1) \\ +k_1 \cdot \Tilde{P}_3^{\rho, R}(n_0-1,n_l-1,n_f,r_0-1,r, d_2-1,k_1-1) + k_2 \cdot \Tilde{P}_3^{\rho, R}(n_0-1,n_l-1,n_f,r_0-1,r, d_2-1,k_2-1) \\+ m_2 \cdot \Tilde{P}_3^{\rho, R}(n_0-1,n_l-1,n_f,r_0-1,r, d_2-1,m_2-1) +l_1 \cdot \Tilde{P}_3^{\rho, R}(n_0-2,n_l-1,n_f,r_0-1,r, d_2-1,l_1-1)\Big) \\+\frac{k_1}{r} \cdot \Big(c_1 \cdot \Tilde{P}_3^{\rho, R}(n_0,n_l-1,n_f,r_0-1,r, c_1-1,k_1-1)+ e_2 \cdot \Tilde{P}_3^{\rho, R}(n_0,n_l-1,n_f,r_0-1,r, e_2-1,k_1-1) \\+d_1 \cdot \Tilde{P}_3^{\rho, R}(n_0-1,n_l-1,n_f,r_0-1,r, d_1-1,k_1-1) \Big)
    +\frac{m_2}{r} \cdot \Big(c_1 \cdot \Tilde{P}_3^{\rho, R}(n_0,n_l-1,n_f,r_0-1,r, c_1-1,m_2-1) \\+ e_2 \cdot \Tilde{P}_3^{\rho, R}(n_0,n_l-1,n_f,r_0-1,r, e_2-1,m_2-1) + d_1 \cdot \Tilde{P}_3^{\rho, R}(n_0-1,n_l-1,n_f,r_0-1,r, d_1-1,m_2-1)\Big) \\
    +\frac{l_1}{r} \cdot \Big(c_1 \cdot \Tilde{P}_3^{\rho, R}(n_0-1,n_l-1,n_f,r_0-1,r, c_1-1,l_1-1) + e_2 \cdot \Tilde{P}_3^{\rho, R}(n_0-1,n_l-1,n_f,r_0-1,r, e_2-1,l_1-1) \\+ d_1 \cdot \Tilde{P}_3^{\rho, R}(n_0-2,n_l-1,n_f,r_0-1,r, d_1-1,l_1-1)\Big).
\end{multline*}
\item The opening of a new recombination loop,
\begin{multline*}
    Recomb=\frac{\rho}{2}\Big((n-n_f-2r_0) \cdot \Tilde{P}_3^{\rho, R}(n_0,n+1,n_f,r_0+1,r+1, a+1,i+1) \\+n_f \cdot \Tilde{P}_3^{\rho, R}(n_0,n+1,n_f-1,r_0+1,r+1, a+1,i+1)\Big).
\end{multline*}
\end{enumerate}

Then the full equation can be expressed as 
\begin{equation*}
    Rate \cdot \Tilde{P}_3^{\rho, R}(n_0,n_l,n_f,r_0,r)=Coal_{NR}+Coal_{R}+Coal_{RR}+Mut_{NR}+Mut_{R}+Recomb.
\end{equation*}

The boundary conditions are
\[
P_3^{\rho, R}(n_0,n_l=1,n_f=0,r_0=0,R, 0,0,0,0,0,0,0,0,0,0,0,0,0,0) = 
\begin{cases}
1 \text{ for } n_f \in \{0, 1, 2\}, \\ 
0 \text{ otherwise}.
\end{cases}
\]

\subsection{Solving the recursions}
The system of equations is then solved iteratively. Due to the multitude of indices involved, the order of implementation needs to be chosen carefully, so that each equation only involved one unknown quantity (to prevent having to use matrix inversion methods). The boundary conditions for $n_l=1$ are
\begin{align*}
    &P_3^{\rho, R}(n_0,n_l=1,n_f=0,r_0=0,r=R, 0,0,0,0,0,0,0,0,0,0,0,0,0,0) = 1, \\
    &P_3^{\rho, R}(n_0,n_l=1,n_f=1,r_0=0,r=R, 0,0,0,0,0,0,0,0,0,0,0,0,0,0) = 1,\\ 
    &P_3^{\rho, R}(n_0,n_l=1,n_f,r_0,r, a,b,c_1,c_2,d_1,d_2,e_1,e_2,i,j,k_1,k_2,l_1,l_2,m_1,m_2) = 0 \;\;\; \text{otherwise},
\end{align*} 
due to restrictions on index range (i.e.\ $r_0<r<R$) or because we require the ARG to have completed $R$ recombination events when it reaches $n_l=1$ lineages. This holds for any $n_0$, and we first solve for $n_0=0$. With $r=R, r_0=0$ all recombination loops required have closed. This forces all subscript indices to be 0 (as these track the states of recombinant edges). We then can solve over $n_l$ and $n_f$ in the same order as for the restriction to the coalescent tree case in Section 2. 

In order to use only data from the boundary conditions we first solve for $n_l=2=n_f$:
\begin{align*}
    (1+ \rho) &\cdot P_3^{\rho, R}(n_0,n_l=2,n_f=2,r_0=0,r=R, 0,0,0,0,0,0,0,0,0,0,0,0,0,0) = \\
    &1 \cdot P_3^{\rho, R}(n_0,n_l=1,n_f=0,r_0=0,r=R, 0,0,0,0,0,0,0,0,0,0,0,0,0,0),
\end{align*}
as this uses only the non-trivial boundary data. Then we can use this information to solve for $n_l=2, n_f=1$:
\begin{align*}
    (1 + \rho + \frac{\theta}{2}) &\cdot P_3^{\rho, R}(n_0,n_l=2,n_f=1,r_0=0,r=R,0,0,0,0,0,0,0,0,0,0,0,0,0,0)= \\
    & 1 \cdot P_3^{\rho, R}(n_0,n_l=1,n_f=0,r_0=0,r=R, 0,0,0,0,0,0,0,0,0,0,0,0,0,0) \\
    &+ \frac{\theta}{2} \cdot P_3^{\rho, R}(n_0,n_l=2,n_f=2,r_0=0,r=R,0,0,0,0,0,0,0,0,0,0,0,0,0,0),
\end{align*}
and then for $n_l=2, n_f=0$. We then increase $n_l$ to 3, and again solve for all $n_f$, working backwards from 3 to 0. In this way we can solve over $n_l$ from 0 to $n$ with each equation only using values previously obtained. 

Now we wish to vary $r$ and $r_0$. As we have all the values for $r=R$, $r_0=0$ we consider equations were $r=R$ still, but now $r_0 = 1$, i.e.\ at each ARG configuration no further recombinations can occur, but one loop is still open. This scenario allows for non-zero values of the subscripts. 
We again start by considering $n_l=2$. As $n_f$ is bounded by $n_l - 2r_0$, this forces $n_f=0$. We then solve the equations for $P_3^{\rho, R}(n_0=2,n_l=2,n_f=0,r_0=1,r=R, a,b,c_1,c_2,d_1,d_2,e_1,e_2,i,j,k_1,k_2,l_1,l_2,m_1,m_2)$ varying each subscript between 0 and 1 in the order $e_2$, $e_1$, $d_2$, $d_1$, $c_2$, $c_1$, $b$, $a$, $e$, $m_2$, $m_1$, $l_2$, $l_1$, $k_2$, $k_1$, $j$, $i$. This ensures that each equation only uses probabilities already calculated, or which are trivially 0 by the restriction $r = a+b+c_1+c_2+d_1+d_2+e_1+e_2 = i+j+k_1+k_2+l_1+l_2+m_1+m_2$. Then we can increase $n_l$ to 3, set $n_f= 3 - 2 = 1$ and again solve over the subscript indices. Then solve for $n_l=3, n_f=0$, and continue in this way increasing $n_l$ up to $n$. 

In this manner we can solve for $r_0 = 2$, still fixing $r=R$, and then iterate $r_0$ up to $R$. This give all the probabilities for $r=R, n_0=0$. The next index to solve over is $r$, here working backwards from the known values of $r=R$ down to $r=0$. Finally, $n_0$ is iterated forwards from 0 to the total number of allowed unresolved edges.

\subsection{Time complexity}

The computation time can be estimated using simple reasoning, despite some of the recursions looking quite complex. If a quantity is recursively defined using $k$ integer arguments, and evaluation of the quantity for fixed values of the $k$ arguments $(m_1,m_2,..,m_k)$ needs evaluation of some function $g(m_1,m_2,..,m_k)$, then two questions need to be considered: how many different arguments are there, and for each, what is $g$? Suppose we have a lower triangular matrix in $k$ dimensions, and for each argument we need to evaluate all smaller arguments, then computation time will grow like $k^2$-th power.  If we only need to refer to arguments smaller by a constant number, then it grows like $k$-th power.

The computation time needed to evaluate these recursions is of the order of $n_0 \cdot n^2 \cdot R^{17}$ where $R$ is the total number of recombinations in the history. This quickly becomes unfeasibly large, but the restriction \[
r = a+b+c_1+c_2+d_1+d_2+e_1+e_2 = i+j+k_1+k_2+l_1+l_2+m_1+m_2,
\] can be exploited to significantly reduce the computation time. If $\mathcal{B}(R)$ is the number of tuples of eight non-negative integers that sum to $R$, then $\mathcal{B}(1,2,3,4,5)= (9,45,165,495,1287)$. Exploiting this gives a reduced computational time of the order of $n_0 \cdot n^2 \cdot \mathcal{B}(R)^2$, a major decrease from $R^{17}$.

\newpage
\subsection{Probability of full ARG recoverability, varying breakpoint}
\phantom{a}

\begin{figure}[h!]
  \centering
    \includegraphics[scale=0.75]{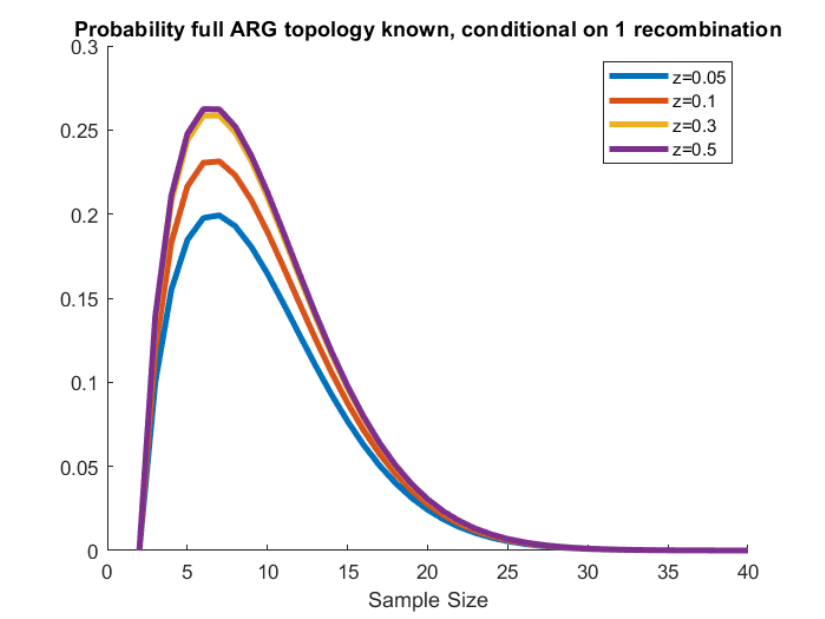}
    \caption{Fixed $\theta=100$, varying breakpoint position $z$ across $[0,0.5]$.}
  \label{fig:one_recomb_vary_z}
 \end{figure} 

\subsection{Detection of a gene conversion} \label{app_gene_conversion}
As observed in the main section, there is a certain amount of flexibility as to the order of the $A$- and $C$-type mutations in the history. Therefore, conditioning on the order of these mutations on both the $\mathcal{E}$ and $\mathcal{F}$ lineages is required. Under the uniform mutation rate assumption, with mutations occurring as competing Poisson processes, the probability of a type $A$ mutation occurring before a type $C$ is $\theta_A/(\theta_A+\theta_C)$, where as before $\theta_A=\theta\, \cdot$ length($A$). Events on distinct lineages are independent.

Due to the breakdown of symmetry, the states for each lineage are given separately in Tables \ref{tab:table3} and \ref{tab:table4}.
\begin{table}[h!]
  \begin{center}
    \caption{States described for recombinant edge $\mathcal{E}$}
    \label{tab:table3}
    \begin{tabular}{l|l} 
      
      State 0 & No coalescence has occurred since the recombination. \\
      \hline
      State 1 & There has been at least one coalescence since the recombination.\\
       & No mutations have occurred since the last coalescence.\\
       \hline
      State 2 & The first of the $A$/$C$-type mutations has occurred\\
       & since the last coalescence. \\
      \hline
      State 3 & The second of the $A$/$C$-type mutations has occurred\\
       & since the last coalescence. This mutation must be \\
       & different to the previous mutation in state 2.\\
      \hline
      State 4 & $\mathcal{E}$ has reached state 3, and undergone one further coalescence.\\
      \hline
      State 5 & Type $B$ mutation has occurred since the last coalescence. \\
    \end{tabular}
  \end{center}
\end{table}
\begin{table}[h!]
  \begin{center}
    \caption{States described for recombinant edge $\mathcal{F}$}
    \label{tab:table4}
    \begin{tabular}{l|l} 
      
      State 0 & No coalescence has occurred since the recombination. \\
      \hline
      State 1 & There has been at least one coalescence since the recombination.\\
       & No mutations have occurred since the last coalescence.\\
       \hline
      State 2 &  A $B$-type mutations has occurred since the last coalescence. \\
      \hline
      State 4 & $\mathcal{F}$ has reached state 3, and undergone one further coalescence.\\
      \hline
      State 5 & The first of the $A$/$C$-type mutations has occurred\\
       & since the last coalescence. \\
      \hline
      State 6 & The second of the $A$/$C$-type mutations has occurred\\
       & since the last coalescence. Again this mutation must be \\
       & different to the previous mutation in state 5.\\
    \end{tabular}
  \end{center}
\end{table}
\break
Note that due to the choice of state labels, $\mathcal{F}$ does not have a State 3 equivalent. Again, we use the phrasing that the ARG being in state $(i,j)$ means $\mathcal{E}$ is in state $i$ and $\mathcal{F}$ is in state $j$.

The recombination will be detectable if $\mathcal{E}$ reaches state 5, or $\mathcal{F}$ reaches state 6, or $\mathcal{E}$ is in a state $>2$ and $\mathcal{F}$ in a state $>1$. If the ARG reaches one of these absorbing states, the subsequent probability of detection is given by $Q^{\rho}(k)$, the probability of no further gene conversion events in the sample. We have $Q^{\rho}(k)= \prod_{m=2}^n (m-1)/(m-1+\rho)$.

Denote the first of the $A$ or $C$ type mutations on lineage $\mathcal{E}$ (resp.~$\mathcal{F}$) as $l_1$ (resp.~$r_1$) and the second as $l_2$ (resp.~$r_2$).

If $l_2 \neq r_2$, the ARG in state $(2,5)$ has the recombination detectable immediately, i.e.\ $P_4^{\rho}(n_l,2,5)=Q^{\rho}(k)$. If $l_2=r_2$, we have the relation 
\begin{equation*}
    \Bigg(\binom{n_l}{2} + \frac{\theta_{l_2}+\theta_{r_2}}{2} +\frac{\rho n_l}{2} \Bigg) P_4^{\rho}(n_l,2,5) =  \Bigg(\binom{n_l}{2}-1\Bigg) P_4^{\rho}(n_l-1,2,5) + \frac{\theta_{r_2}}{2} Q^{\rho}(k) +\frac{\theta_{l_2}}{2} Q^{\rho}(k),
\end{equation*}
and for every combination of $l_i,r_i$:
\begin{equation*}
    \Bigg(\binom{n_l}{2} + \frac{\theta_{l_2}+\theta_{r_1}}{2} +\frac{\rho n_l}{2} \Bigg) P_4^{\rho}(n_l,2,4) =  \Bigg(\binom{n_l}{2}-1\Bigg) P_4^{\rho}(n_l-1,2,4) + \frac{\theta_{r_1}}{2} P_4^{\rho}(n_l,2,5) +\frac{\theta_{l_2}}{2} Q^{\rho}(k)
\end{equation*}
\begin{equation*}
    \Bigg(\binom{n_l}{2} + \frac{\theta_{l_2}}{2}  +\frac{\rho n_l}{2} \Bigg) P_4^{\rho}(n_l,2,2) =  \binom{n-1}{2} P_4^{\rho}(2,2) + (n-2) P_4^{\rho}(n_l-1,2,4) +\frac{\theta_{l_2}}{2} Q^{\rho}(k)
\end{equation*}
\begin{equation*}
    \Bigg(\binom{n_l}{2} + \frac{\theta_{l_1}+\theta_{r_2}}{2} +\frac{\rho n_l}{2} \Bigg) P_4^{\rho}(n_l,1,5) =  \Bigg(\binom{n_l}{2}-1\Bigg) P_4^{\rho}(n_l-1,1,5) + \frac{\theta_{r_2}}{2} Q^{\rho}(k) +\frac{\theta_{l_1}}{2} P_4^{\rho}(n_l,2,5)
\end{equation*}
\begin{equation*}
    \Bigg(\binom{n_l}{2} + \theta_{B} +\frac{\rho n_l}{2} \Bigg) P_4^{\rho}(n_l,4,1) =  \Bigg(\binom{n_l}{2}-1\Bigg) P_4^{\rho}(n_l-1,4,1) + \theta_{B}Q^{\rho}(k)
\end{equation*}
\begin{equation*}
    \Bigg(\binom{n_l}{2} + \frac{\theta_{l_1}+\theta_{r_1}}{2} +\frac{\rho n_l}{2} \Bigg) P_4^{\rho}(n_l,1,4) =  \Bigg(\binom{n_l}{2}-1\Bigg) P_4^{\rho}(n_l-1,1,4) + \frac{\theta_{r_1}}{2} P_4^{\rho}(n_l,1,5) +\frac{\theta_{l_1}}{2} Q^{\rho}(k)
\end{equation*}
\begin{equation*}
    \Bigg(\binom{n_l}{2} + \frac{\theta_{B}}{2}  +\frac{\rho n_l}{2} \Bigg) P_4^{\rho}(n_l,3,1) =  \binom{n-1}{2} P_4^{\rho}(n_l-1,3,1) + (n-2) P_4^{\rho}(n_l-1,4,1) +\frac{\theta_{B}}{2} Q^{\rho}(k)
\end{equation*}
\begin{equation*}
    \Bigg(\binom{n_l}{2} + \frac{\theta_{l_1}}{2}  +\frac{\rho n_l}{2} \Bigg) P_4^{\rho}(n_l,1,2) =  \binom{n-1}{2} P_4^{\rho}(n_l-1,1,2) + (n-2) P_4^{\rho}(n_l-1,1,4) +\frac{\theta_{l_1}}{2} P_4^{\rho}(n_l,2,2)
\end{equation*}
\begin{equation*}
    \Bigg(\binom{n_l}{2} + \frac{\theta_{B}+\theta_{r_2}}{2} +\frac{\rho n_l}{2} \Bigg) P_4^{\rho}(n_l,2,1) =  \Bigg(\binom{n_l}{2}-1\Bigg) P_4^{\rho}(n_l-1,2,1) + \frac{\theta_{r_2}}{2} P_4^{\rho}(n_l,3,1) +\frac{\theta_{B}}{2} P_4^{\rho}(n_l,2,2)
\end{equation*}
\begin{equation*}
    \Bigg(\binom{n_l}{2} + \frac{\theta_{B}+\theta_{l_1}}{2} +\frac{\rho n_l}{2} \Bigg) P_4^{\rho}(n_l,1,1) =  \Bigg(\binom{n_l}{2}-1\Bigg) P_4^{\rho}(n_l-1,1,1) + \frac{\theta_{l_1}}{2} P_4^{\rho}(n_l,2,1) +\frac{\theta_{B}}{2} P_4^{\rho}(n_l,1,2)
\end{equation*}
\begin{equation*}
    \Bigg(\binom{n_l}{2} + \frac{\theta_{r_2}}{2}  +\frac{\rho n_l}{2} \Bigg) P_4^{\rho}(n_l,0,5) =  \binom{n-1}{2} P_4^{\rho}(n_l-1,0,5) + (n-2) P_4^{\rho}(n_l-1,1,5) +\frac{\theta_{r_2}}{2} Q^{\rho}(k)
\end{equation*}
\begin{equation*}
    \Bigg(\binom{n_l}{2} + \frac{\theta_{r_1}}{2}  +\frac{\rho n_l}{2} \Bigg) P_4^{\rho}(n_l,0,4) =  \binom{n-1}{2} P_4^{\rho}(n_l-1,0,4) + (n-2) P_4^{\rho}(n_l-1,1,4) +\frac{\theta_{r_1}}{2} P_4^{\rho}(n_l,0,5)
\end{equation*}
\begin{equation*}
    \Bigg(\binom{n_l}{2} + \frac{\theta_{B}}{2}  +\frac{\rho n_l}{2} \Bigg) P_4^{\rho}(n_l,4,0) =  \binom{n-1}{2} P_4^{\rho}(n_l-1,4,0) + (n-2) P_4^{\rho}(n_l-1,4,1) +\frac{\theta_{B}}{2} Q^{\rho}(k)
\end{equation*}
\begin{equation*}
    \Bigg(\binom{n_l}{2} + \frac{\rho n_l}{2} \Bigg) P_4^{\rho}(n_l,3,0) =  \binom{n-2}{2} P_4^{\rho}(n_l-1,3,0) + (n-2) (P_4^{\rho}(n_l-1,4,0) +P_4^{\rho}(n_l-1,3,1))
\end{equation*}
\begin{equation*}
    \Bigg(\binom{n_l}{2} + \frac{\rho n_l}{2} \Bigg) P_4^{\rho}(n_l,0,2) =  \binom{n-2}{2} P_4^{\rho}(n_l-1,0,2) + (n-2) (P_4^{\rho}(n_l-1,0,4) +P_4^{\rho}(n_l-1,1,2))
\end{equation*}
\begin{equation*}
    \Bigg(\binom{n_l}{2} + \frac{\theta_{l_2}}{2}  +\frac{\rho n_l}{2} \Bigg) P_4^{\rho}(n_l,2,0) =  \binom{n-1}{2} P_4^{\rho}(n_l-1,2,0) + (n-2) P_4^{\rho}(n_l-1,2,1) +\frac{\theta_{l_2}}{2} P_4^{\rho}(n_l,3,0)
\end{equation*}
\begin{equation*}
    \Bigg(\binom{n_l}{2} + \frac{\theta_{l_1}}{2}  +\frac{\rho n_l}{2} \Bigg) P_4^{\rho}(n_l,0,1) =  \binom{n-1}{2} P_4^{\rho}(n_l-1,0,1) + (n-2) P_4^{\rho}(n_l-1,1,1) +\frac{\theta_{l_1}}{2} P_4^{\rho}(n_l,0,2)
\end{equation*}
\begin{equation*}
    \Bigg(\binom{n_l}{2} + \frac{\theta_{B}}{2}  +\frac{\rho n_l}{2} \Bigg) P_4^{\rho}(n_l,1,0) =  \binom{n-1}{2} P_4^{\rho}(n_l-1,1,0) + (n-2) P_4^{\rho}(n_l-1,1,1) +\frac{\theta_{B}}{2} P_4^{\rho}(n_l,2,0)
\end{equation*}
\begin{equation*}
    \Bigg(\binom{n_l}{2} + \frac{\rho n_l}{2} \Bigg) P_4^{\rho}(0,0) =  \binom{n-2}{2} P_4^{\rho}(n_l-1,0,0) + (n-2) (P_4^{\rho}(n_l-1,0,1) +P_4^{\rho}(n_l-1,1,0))
\end{equation*}

\end{document}